\documentclass[11pt,a4paper]{article}
\pdfoutput=1

\usepackage{jheppub}

\usepackage{graphicx,amssymb,amsmath,mathtools,bm,bbm}
\usepackage{hepunits}
\usepackage{physics}
\usepackage{epsfig}

\usepackage{hyperref}
\usepackage{url}
\hypersetup{pdfnewwindow=true}

\usepackage{tabularx}
\newcolumntype{C}[1]{>{\hsize=#1\hsize\centering\arraybackslash}X}%

\usepackage[utf8]{inputenc}

\def\ep{\epsilon}
\def\sigh{\hat{\sigma}}
\def\sigt{\tilde{\sigma}}
\def\la{\langle}
\def\ra{\rangle}
\def\R{\mathrm{Re}}
\def\mR{\mu^2_R}
\def\mF{\mu^2_F}
\def\cm{\mathcal{M}}
\def\cf{\mathcal{F}}
\def\as{{\alpha_s}}
\def\msbar{$\overline{\mathrm{MS}}$ }

\def\nn{\nonumber}
\def\be{\begin{equation}}
\def\ee{\end{equation}}
\newcommand\f[2]{\frac{#1}{#2}}

\allowdisplaybreaks

\preprint{
  \begin{flushright}
    Cavendish-HEP-19/13 \\
    IFJPAN-IV-2019-11 \\
    P3H-19-024 \\
    TTK-19-29
  \end{flushright}}

\title{Single-jet inclusive rates with exact color at
  $\mathcal{O}(\alpha_s^4)$}

\author[a]{Micha\l\ Czakon,}
\emailAdd{mczakon@physik.rwth-aachen.de}
\author[b]{Andreas van Hameren,}
\emailAdd{Andre.Hameren@ifj.edu.pl}
\author[c]{Alexander Mitov}
\emailAdd{adm74@cam.ac.uk}
\author[c]{and Rene Poncelet}
\emailAdd{poncelet@hep.phy.cam.ac.uk}

\affiliation[a]{Institut f\"{u}r Theoretische Teilchenphysik und
  Kosmologie,
  RWTH Aachen University, \\
  D-52056 Aachen, Germany}

\affiliation[b]{Institute of Nuclear Physics,
  Polish Academy of Sciences, \\
  ul. Radzikowskiego 152, 31-342, Cracow, Poland}

\affiliation[c]{Cavendish Laboratory,
  University of Cambridge, \\
  CB3 0HE Cambridge, UK}

\abstract{Next-to-next-to-leading order QCD predictions for single-,
  double- and even triple-differential distributions of jet events in
  proton-proton collisions have recently been obtained using the
  \textsc{NNLOjet} framework based on antenna subtraction. These results
  are an important input for Parton Distribution Function fits to
  hadron-collider data. While these calculations include all of the
  partonic channels occurring at this order of the perturbative
  expansion, they are based on the leading-color approximation in the
  case of channels involving quarks and are only exact in color in the
  pure-gluon channel. In the present publication, we verify that the
  sub-leading color effects in the single-jet inclusive
  double-differential cross sections are indeed negligible as far as
  phenomenological applications are concerned. This is the first
  independent and complete calculation for this observable. We also
  take the opportunity to discuss the necessary modifications of the
  sector-improved residue subtraction scheme that made this work
  possible.}

\begin{document}

\maketitle

\newpage


\section{Introduction}

Pure jet observables are not only interesting within the portfolio of
Standard Model (SM) measurements but also as tools for New Physics
searches. When paired with high-precision predictions within Quantum
Chromodynamics (QCD), they are an important input for Parton
Distribution Function (PDF) fits. Consistency of the latter
application requires theoretical predictions at the same order of
perturbation theory as the specified accuracy of the given PDF set. In
consequence, it is nowadays indispensable to determine differential
distributions of jet events at the next-to-next-to-leading order
(NNLO) in QCD. Although not the subject of the present publication,
non-perturbative effects from underlying event and hadronisation
\cite{Khachatryan:2016wdh} as well as electroweak corrections
\cite{Dittmaier:2012kx} should also be included for a realistic
comparison with measurement data.

A calculation of jet rates at NNLO in QCD remains a very challenging
project. Until now, results could only have been obtained with the
help of the antenna subtraction scheme \cite{GehrmannDeRidder:2005cm,
GehrmannDeRidder:2005hi, GehrmannDeRidder:2005aw, Daleo:2006xa,
Daleo:2009yj, Boughezal:2010mc, Gehrmann:2011wi,
GehrmannDeRidder:2012ja, Currie:2013vh, Bernreuther:2011jt,
Bernreuther:2013uma, Abelof:2011jv, Abelof:2012he, Abelof:2014fza} as
implemented in the \textsc{NNLOjet} framework
\cite{Currie:2018oxh}. While early studies concentrated on the
pure-gluon case \cite{Currie:2013dwa, Ridder:2013mf}, several
publications \cite{Currie:2017eqf, Currie:2016bfm,
Gehrmann-DeRidder:2019ibf, Currie:2018xkj} appeared recently that
present results including the complete set of partonic channels. It
turns out that quark--gluon and quark--quark scattering dominates jet
rates at high transverse momenta and/or rapidities. These
contributions are, therefore, necessary for a complete description. On
the other hand, the current \textsc{NNLOjet} implementation is based
on the leading-color approximation for all channels but the pure-gluon
channel. It has been argued in the past that inclusion of the
sub-leading color effects at $\mathcal{O}(\alpha_S^4)$ while keeping
exact color dependence for the lower order contributions will have a
negligible impact on the predictions. Nevertheless, it is necessary to
verify this statement by explicit calculation. This is the main
purpose of the present publication.

The published results on jet rates correspond to several different
setups. In particular, \cite{Currie:2017eqf, Currie:2016bfm}
correspond to 7 TeV ATLAS data, while \cite{Gehrmann-DeRidder:2019ibf}
to 8 TeV and \cite{Currie:2018xkj} to 13 TeV CMS data. Single-jet
inclusive cross sections (every identified jet in an event is
accounted for in the histogram) that are double-differential in the
jet transverse momentum, $p_T$, and rapidity, $|y|$, have been
published for 7 and 13 TeV Large Hadron Collider (LHC) center-of-mass
energies and various jet radii, $R$, defined with the anti-$k_T$ jet
algorithm \cite{Cacciari:2008gp}. Furthermore, for the same
center-of-mass energies, there are available results for di-jet cross
sections that are double differential in the di-jet invariant mass,
$m_{jj}$, and the jet rapidity difference, $|y^*| \equiv |y_1 -
y_2|/2$. Finally, there is a published result for di-jet cross
sections corresponding to 8 TeV CMS data that is triple-differential
in the average transverse momentum, $p_{T,\mathrm{avg}} \equiv
(p_{T,1} + p_{T,2})/2$, rapidity difference, $|y^*|$, and the boost,
$y_b \equiv |y_1 + y_2|/2$. For convenience of the reader, we
summarise the available results in Appendix~\ref{app:overview}.

Di-jet cross sections have consistently been evaluated with central
renormalisation and factorisation scales $\mu_R = \mu_F = \mu =
m_{jj}$. However, it turned out that the scale choice for single-jet
inclusive cross sections is a non-trivial issue, since seemingly
well-justified choices lead to large differences in the predictions as
illustrated by the results for 7 TeV center-of-mass energy using the
customary scales $\mu = p_T$ (each jet input into the histogram with
the cross section evaluated with the scale equal to that particular
jet transverse momentum) and $\mu = p_{T,1}$ (cross section scale set
to the hardest-jet transverse momentum). The question of scale setting
has been very thoroughly studied in \cite{Currie:2018xkj}. The final
recommendation of that publication is to use either the jet-based
scale $\mu = 2p_T$ or the event-based scale $\mu = \hat{H}_T$ (the
scalar sum of the transverse momenta of the partons in the event).

The high computational cost of the calculation of jet rates with our
software (see Section~\ref{sec:results}) enabled us to only perform
one complete Monte Carlo simulation for the present publication. Since
the setups of the different theoretical predictions described above
differ in energy, jet transverse momentum cuts and jet radii, we had
to choose a single specific setup to study the sub-leading color
effects. Our goal was to compare results for a classic observable used
in PDF fits that has been previously evaluated with one of the
recommended scales. In view of these considerations, we have chosen to
evaluate the single-jet inclusive cross section for 13 TeV
center-of-mass energy available from \cite{Currie:2018xkj} (jet radius
$R=0.7$) using the jet-based scale $\mu = 2p_T$. We should also point
out that there are no available numerical values for cross sections or
cross sections ratios in any of the publications
\cite{Currie:2017eqf, Currie:2016bfm, Gehrmann-DeRidder:2019ibf,
Currie:2018xkj} - only histograms have been
provided. Appendix~\ref{app:k-factors} contains our results for the
K-factors, i.e.\ ratios of cross sections evaluated at NNLO and NLO in
QCD with the same PDF set. This is a first step towards an easy
inclusion of jet data in NNLO PDF fits. In the future, it would
certainly be desirable to provide \textsc{fastNLO} \cite{Kluge:2006xs,
Wobisch:2011ij, Britzger:2012bs} or \textsc{APPLGRID}
\cite{Carli:2010rw} tables for all setups. We intend to undertake this
task once more computational resources become available.

Besides the study of sub-leading color effects in jet rates, there is
another aspect to the present publication. The calculation of a cross
section at NNLO in QCD requires a method to handle infrared (IR)
singularities occurring in contributions of different final state
multiplicities. Apart from the already mentioned antenna subtraction
scheme, there are several other methods currently being developed for
this purpose:
the \textsc{CoLoRfulNNLO} scheme
\cite{Somogyi:2005xz, Somogyi:2006da, Somogyi:2006db, Aglietti:2008fe,
  DelDuca:2013kw},
$q_T$-slicing
\cite{Catani:2007vq, Bonciani:2015sha},
$N$-jettiness slicing
\cite{Gaunt:2015pea, Boughezal:2015dva, Boughezal:2016zws,
  Moult:2016fqy, Moult:2017jsg, Ebert:2018lzn, Boughezal:2018mvf,
  Boughezal:2019ggi},
sector-improved residue subtraction
\cite{Czakon:2010td, Czakon:2011ve, Czakon:2014oma}
and its spin-off called nested soft-collinear subtraction scheme
\cite{Caola:2017dug, Caola:2019nzf, Caola:2019pfz},
the projection-to-Born method
\cite{Cacciari:2015jma,Currie:2018fgr},
local analytic sector subtraction
\cite{Magnea:2018hab}
and geometric IR subtraction
\cite{Herzog:2018ily}.
The results that we report in Section~\ref{sec:results} have been
obtained with our implementation of the sector-improved residue
subtraction in the C++ library \textsc{Stripper} (SecToR Improved
Phase sPacE for real Radiation). However, we have introduced several
improvements w.r.t.\ to Ref.~\cite{Czakon:2014oma}. These improvements
imply for instance a minimal number of subtraction terms per phase
space point. They also require a modified reduction of the
construction in Conventional Dimensional Regularisation (CDR) to four
dimensions ('t Hooft-Veltman scheme). In the present publication, we
discuss these issues in detail.

The paper is organised as follows. In the next section, we discuss
modifications of the sector-improved residue subtraction. These
consist of an improved phase space parameterisation and a new approach
to the dimensional reduction of the formulation of the scheme. The
section is closed with details on the implementation and
tests. Subsequently, we present our results for the single-jet
inclusive cross sections at 13 TeV. The main text is closed with an
outlook. Appendices provide an overview of published results on jet
cross sections, define the notation for cross section contributions,
provide a list of expressions necessary for the implementation of the
subtraction scheme in four dimensions, and, finally, provide numerical
values of the NNLO K-factors.


\section{Minimal sector-improved residue subtraction}

We define a subtraction scheme to be minimal, if it has the minimal
number of subtraction terms (defined by final state kinematics) for a
given phase space point. Consider a phase space configuration with
$n+2$ final state particles contributing to a cross section at
next-to-next-to-leading order of perturbation theory. A configuration
corresponding to a single soft (one gluon with vanishing energy) or
collinear (two partons collinear) limit will have $n+1$ resolved final
state particles (single-unresolved configuration), while a
configuration corresponding to a double-soft (two gluons or a
quark-anti-quark pair with vanishing energy) or triple-collinear
(three partons collinear) limit will have $n$ resolved final state
particles (double-unresolved configuration). Let us divide the phase
space into sectors according to collinear limits (see Section 3 of
Ref.~\cite{Czakon:2014oma} for more details). A sector that only
allows for one singular configuration (divergent cross section
contribution) in a specific single collinear limit is called
single-collinear. A sector that allows for one singular configuration
in a specific limit of two pairs of collinear partons is called
double-collinear. Finally, a sector that allows for one singular
configuration in a specific limit of three collinear partons is called
triple-collinear. It is easy to convince one-self that the minimal
number of subtraction terms in a single-collinear sector is
one. Similarly, the minimal number of subtraction terms in a
double-collinear sector is three (two for single-unresolved
configurations and one for the double-unresolved
configuration). Finally, the minimal number of subtraction terms in a
triple-collinear sector is four (three for single-unresolved
configurations and one for the double-unresolved configuration). The
phase space construction of the sector-improved residue subtraction
scheme as defined in Ref.~\cite{Czakon:2014oma} generates more
subtraction terms. Here, we present an alternative parameterization
of the phase space that,
due to the additional sector decomposition in the triple-collinear
sector, never requires more than three subtraction terms for a given
phase space point. It turns out, however, that the four-dimensional
formulation of the scheme requires modifications with this phase
space. On the other hand, the methods presented here allow for
numerical checks of pole cancellation in CDR for a fixed Born phase
space point \cite{Behring:2018cvx, PhDthesisPoncelet}. In contrast, an
implementation of Ref.~\cite{Czakon:2014oma} yields finite results at
the level of distributions only.

We note, finally, that a single subtraction configuration in
single-collinear parameterisations has also been obtained in the FKS
subtraction scheme \cite{Frixione:1995ms} implementations of
Refs.~\cite{Frixione:2002ik, Frixione:2007vw}. The approach presented
here is conceptually different, and allows to cover next-to-leading
and next-to-next-leading order cases on the same footing. It might be
extensible to even higher orders.

\subsection{Modified phase space parameterisations}
\label{sec:PhaseSpace}


\subsubsection{Phase space mapping}
\label{sec:PhaseSpaceMapping}

Let us introduce the following notation for the phase space measure
corresponding to a single particle with mass $m \geq 0$ and momentum
$k$ in $d$ dimensions
\begin{equation}
\mathrm{d} \mu_m(k) \equiv \frac{\mathrm{d}^d k}{(2\pi)^d} \, 2\pi
\delta\big(k^2 - m^2\big) \theta\big(k^0\big) \equiv
\frac{\mathrm{d}^d k}{(2\pi)^d} \, 2\pi \delta_+\big(k^2 - m^2\big) \;
.
\end{equation}
The complete phase space for a process involving $n_q \neq
0$ \footnote{In the special case of only four massless partons in the
final state, all corresponding to unresolved and reference momenta,
i.e. $n_q = 0$, the parameterisation of Ref.~\cite{Czakon:2014oma}
section 4.3.2 already satisfies our requirements: there is only one
double- and two single-unresolved configurations.} (not explicitly
parameterised) resolved momenta $q_i$, $n_u = 1,2$ unresolved momenta
$u_i = u_i^0 \hat{u}_i$, and $0 < n_r \leq n_u$ reference momenta $r_i$,
$n_{fr}$ of which are in the final state, can be decomposed as follows
\begin{equation}
\begin{split}
\mathrm{d}\bm{\Phi}_n &= \prod_{i=1}^{n_q} \mathrm{d} \mu_{m_i}(q_i)
\prod_{j=1}^{n_{fr}} \mathrm{d} \mu_0(r_j) \prod_{k=1}^{n_u}
\mathrm{d} \mu_0(u_k) \, (2\pi)^d \delta^{(d)}\Big(\sum_{i=1}^{n_q}
q_i + \sum_{j=1}^{n_{fr}} r_j + \sum_{k=1}^{n_u} u_k - P\Big)
\\[0.2cm] &= \frac{\mathrm{d} Q^2}{2\pi} \, \mathrm{d} \mu_Q(q)
\prod_{j=1}^{n_{fr}} \mathrm{d} \mu_0(r_j) \prod_{k=1}^{n_u}
\mathrm{d} \mu_0(u_k) \, (2\pi)^d \delta^{(d)}\Big(q +
\sum_{j=1}^{n_{fr}} r_j + \sum_{k=1}^{n_u} u_k - P\Big) \\[0.2cm]
& \qquad \times \prod_{i=1}^{n_q} \mathrm{d} \mu_{m_i}(q_i) \,
(2\pi)^d \delta^{(d)}\Big(\sum_{i=1}^{n_q} q_i - q\Big) \\[0.2cm] &=
\mathrm{d} Q^2 \Bigg[ \prod_{j=1}^{n_{fr}} \mathrm{d} \mu_0(r_j)
\prod_{k=1}^{n_u} \mathrm{d} \mu_0(u_k) \, \delta_+\Big(\Big( P -
\sum_{j=1}^{n_{fr}} r_j - \sum_{k=1}^{n_u} u_k \Big)^2 - Q^2\Big)
\Bigg] \\[0.2cm] & \qquad \times \prod_{i=1}^{n_q} \mathrm{d}
\mu_{m_i}(q_i) \, (2\pi)^d \delta^{(d)}\Big(\sum_{i=1}^{n_q} q_i -
q\Big) \; ,
\end{split}
\end{equation}
where $n = n_q + n_{fr} + n_u$ and $P$ is the total initial state
momentum. In the second line, we have inserted an intermediate
momentum $q$ with invariant mass $Q \geq \sum_{i=1}^{n_q} m_i$. In the
third line, however, we have performed the integration over $q$,
leaving an integration measure in the square brackets, which only
depends on the fixed invariant mass of $q$.

The reference and unresolved momenta are the momenta that are allowed
to correspond to a singular configuration in a given sector. Reference
momenta and unresolved momenta have different behaviour w.r.t.\ the
soft limit: a soft reference momentum does not generate a singularity.
In case of a single- and triple-collinear sector one reference momentum
$r$ is needed, while in the double-collinear sectors two reference momenta
$r_1$ and $r_2$ are required.

We now introduce a mapping from the full phase space to the Born phase
space
\begin{equation}
\{ P, r_j, u_k \} \, \longrightarrow \, \{ \tilde{P}, \tilde{r}_j \} \; ,
\end{equation}
which is invertible for fixed unresolved momenta
\begin{equation}
\{ \tilde{P}, \tilde{r}_j, u_k \} \, \longrightarrow \, \{ P, r_j, u_k \} \; ,
\end{equation}
and which conserves the invariant mass of the intermediate state $q$
\begin{equation}
\label{eq:qconstraint}
\tilde{q}^2 = q^2 \; , \quad \tilde{q} = \tilde{P} - \sum_{j=1}^{n_{fr}} \tilde{r}_j \; .
\end{equation}
{\bf The mapping only involves a rescaling of the reference
  momenta}. Specifically, for a final state reference momentum we
require
\begin{equation}
r = x \, \tilde{r} \; ,
\end{equation}
where $x$ is given by a function $f_x$ of the full kinematics, in
particular of $r$. The phase space measure is modified
\begin{equation}
\begin{split}
\mathrm{d} \mu_0(r) &= \mathrm{d} \mu_0(r) \, \mathrm{d}x \,
\delta\big(x - f_x(r)\big) \, \mathrm{d}^d \tilde{r} \,
\delta^{(d)}(\tilde{r} - r/x) \\[0.2cm] &= \mathrm{d} \mu_0(\tilde{r})
\, \mathrm{d}x \, \theta(x) \, x^{d-2} \, \delta\big(x - f_x(x \,
\tilde{r})\big) \\[0.2cm] &= \mathrm{d} \mu_0(\tilde{r}) \, \theta(x)
\, x^{d-3} \Bigg[ - \frac{\partial}{\partial x} \frac{f_x(x \,
\tilde{r})}{x} \Bigg]^{-1} \Bigg|_{x = f_x(x \, \tilde{r})} \; .
\end{split}
\end{equation}
For an initial state reference, we require similarly
\begin{equation}
r = \tilde{r} / z \; ,
\end{equation}
where $z$ is given by a function $f_z$ of the full kinematics, in
particular of $r$. We consider the phase space measure together with
the integration over the parton momentum fraction $x$ with the parton
distribution function $\phi$
\begin{equation}
\begin{split}
\mathrm{d}x \, \phi(x) &= \mathrm{d}x \, \phi(x) \,
\mathrm{d}\tilde{x} \, \delta\big(\tilde{x} - f_z(x \, p_h) \, x\big)
\\[0.2cm] &= \mathrm{d}\tilde{x} \, \phi(\tilde{x} / z) \, \theta(z -
\tilde{x}) \Bigg[ -z^2 \frac{\partial}{\partial z}
\frac{f(\tilde{r}/z)}{z} \Bigg]^{-1} \Bigg|_{z = f(\tilde{r}/z)} \; ,
\end{split}
\end{equation}
where $p_h$ is the initial state hadron momentum.

The Born configuration is only properly defined, i.e. physical, if the
rescaling parameters are always non-negative and $\tilde{q}^0 \geq 0$.
The phase space measure may be rewritten as
\begin{equation}
\begin{split}
\mathrm{d}\bm{\Phi}_n &= \mathrm{d} Q^2 \Bigg[ \prod_{j=1}^{n_{fr}}
\mathrm{d} \mu_0(\tilde{r}_j) \, \delta_+\Big(\Big( \tilde{P} -
\sum_{j=1}^{n_{fr}} \tilde{r}_j \Big)^2 - Q^2\Big) \prod_{k=1}^{n_u}
\mathrm{d} \mu_0(u_k) \, \theta\big(\{ u_l \} \in \mathcal{U}\big) \,
\mathcal{J} \Bigg] \\[0.2cm] &\qquad \times \prod_{i=1}^{n_q}
\mathrm{d} \mu_{m_i}(q_i) \, (2\pi)^d
\delta^{(d)}\Big(\sum_{i=1}^{n_q} q_i - \tilde q\Big) \; ,
\end{split}
\end{equation}
where $\theta\big(\{ u_l \} \in \mathcal{U}\big)$ represents
constraints on the unresolved momenta, and $\mathcal{J}$ is the
Jacobian of the transformation. Furthermore, we have used
Eq.~\eqref{eq:qconstraint} and the implied existence of a Lorentz
transformation $\Lambda$, $q = \Lambda \, \tilde{q}$, together with
the Lorentz invariance of $\mathrm{d} \mu_{m_i}(q_i)$. The Born phase
space measure is clearly singled out, which leads to the following
algorithm for the construction of the full phase space
\begin{enumerate}
\item generate a Born configuration;
\item generate unresolved momenta subject to constraints;
\item determine the rescaling parameters and, by the same, the full
  reference momenta;
\item determine the Lorentz transformation yielding $q$ from
  $\tilde{q}$,  and apply it to the final state momenta of the Born
  configuration;
\item multiply the weight by the Jacobian.
\end{enumerate}
One advantage of this procedure is that it allows for the use of a
multi-channel phase space generator for the Born configuration, which
is particularly useful in case of intermediate
resonances. Furthermore, electroweak decays should not affect
efficiency, since intermediate invariant masses are not
modified in the absence of QCD radiation from the decay products.

The rescaling parameters are fixed by Eq.~\eqref{eq:qconstraint} and
an additional constraint in the case of two references (see
section~\ref{sec:TwoRefMom}).  The relations always involve a sum (final state
reference) or a difference (initial state reference) of a reference
momentum and an unresolved momentum. Thus, if $u = \alpha \, r$
(collinear or soft limit), for some unresolved, $u$, and reference,
$r$, momenta, then the constraints fix $r \pm u$ (the observable or
initial state momentum) independently of $u$. This is the reason for
the {\bf minimal number of subtraction kinematics} in all cases but
the single-unresolved configurations of the triple-collinear
parameterisation (see Section \ref{sec:SUEnergies}).

In the case of initial state references, the energy and rapidity of
the initial state is modified. In order to have a minimal number of
configurations, it is necessary to choose a frame with constant boost
w.r.t.\ the laboratory, e.g.\ the laboratory frame or the center-of-mass
frame of the underlying Born configuration. On the other hand, the
center-of-mass frame of the $(n+n_u)$-configuration cannot be used as
it does not satisfy this constraint.

The same $(n+n_u)$-configuration corresponds to different Born
configurations for different parameterisations (single-collinear,
triple-collinear and double-collinear). Thus, if the angles and
energies of the unresolved momenta are required to have the same
meaning across parameterisations, it is not possible to use the
center-of-mass frame of the Born configuration either. This is
important, since using the same angle and energy definition for all
configurations yields simpler 't Hooft-Veltman corrections. For this
reason {\bf we use the laboratory system} in the construction of the
phase space.

\subsubsection{One reference momentum}
\label{sec:OneRefMom}

We explicitly treat the triple-collinear parameterisation. The
single-collinear case is recovered by setting $u_2 = 0$. The following
constraints determine $x$ or $z$
\begin{alignat*}{2}
& \text{\bf final state reference:} & \hspace{1cm} &\text{\bf initial state
  reference:} \nonumber
\\[0.2cm]
&\big( P - r - u_1 - u_2 \big)^2 = \big( P - \tilde{r} \big)^2 \; ,
&&\big( r + p - u_1 - u_2 \big)^2 = \big( \tilde{r} + p \big)^2 \; ,
\\[0.2cm]
&x = \frac{P \cdot r}{(P - r) \cdot (r + u_1 + u_2) - u_1 \cdot u_2} \; ,
&&z = \frac{(r + p) \cdot (r - u_1 - u_2) + u_1 \cdot u_2}{p \cdot r} \; ,
\\[0.2cm]
&x = \frac{P \cdot (\tilde{r} - u_1 - u_2) + u_1 \cdot u_2}{(P - u_1 -
  u_2) \cdot \tilde{r}} \; ,
&&z = \frac{(p - u_1 - u_2) \cdot \tilde{r}}{p \cdot (\tilde{r} + u_1 +
  u_2 ) - u_1 \cdot u_2} \; ,
\\[0.2cm]
&\big( u_1^0 \big)_{\mathrm{\max}} = \frac{P \cdot \tilde{r}}{P \cdot
  \hat{u}_1} \; ,
&&\big( u_1^0 \big)_{\mathrm{\max}} = (1 - \tilde{x}) \frac{p \cdot
  \tilde{r}/\tilde{x}}{(\tilde{r}/\tilde{x} + p) \cdot \hat{u}_1} \; ,
\\[0.2cm]
&\big( u_2^0 \big)_{\mathrm{\max}} = \frac{P \cdot (\tilde{r} -
  u_1)}{(P - u_1) \cdot \hat{u}_2} \; ,
&&\big( u_2^0 \big)_{\mathrm{\max}} = \frac{(\tilde{r}/\tilde{x} + p)
  \cdot \big( (1 - \tilde{x}) \, \tilde{r}/\tilde{x} -
  u_1\big)}{(\tilde{r}/\tilde{x} + p - u_1) \cdot \hat{u}_2} \; ,
\\[0.2cm]
&\mathcal{J} = \frac{x^{d-3} \, P \cdot \tilde{r}}{(P - u_1 - u_2)
  \cdot \tilde{r}} \; ,
&&\mathcal{J} = \frac{p \cdot \tilde{r}}{(p - u_1 - u_2) \cdot
  \tilde{r}} \; .
\end{alignat*}
They have the following properties
\begin{enumerate}

\item the Born configuration is well defined for any full
  configuration, i.e. $x, z, \tilde{q}^0 \geq 0$, and $x, z \leq 1$,

\item $x$, $z$ are monotonically decreasing functions of $u_i^0$, if
  $u_j$ with $i \neq j$ is fixed.

\end{enumerate}
In consequence, an iterative energy parameterisation with $u_1^0$
determined in the range $\big[ 0, \, \big( u_1^0
\big)_{\mathrm{max}} \big]$, followed by $u_2^0$ in the range $\big[
0, \, \big( u_2^0 \big)_{\mathrm{max}}(u_1) \big]$ covers the full
phase space. The maxima of the energies, $\big( u_i^0
\big)_{\mathrm{max}}$, are obtained at $x = 0$ or $z =
\tilde{x}$. Due to the further requirement $u_1^0 \geq u_2^0$
necessary to factorise double-soft limits, we introduce the
parameterisation
\begin{equation}
\label{eq:OneRefMomEnergies}
u_1^0 = \big( u_1^0 \big)_{\mathrm{max}} \, \xi_1 \; , \quad
u_2^0 = \big( u_1^0 \big)_{\mathrm{max}} \, \xi_1 \xi_2 \underbrace{\min \Bigg[
1, \, \frac{1}{\xi_1} \frac{\big( u_2^0 \big)_{\mathrm{max}}}{\big(
u_1^0 \big)_{\mathrm{max}}} \Bigg]}_{\equiv \bar{\xi}_2} \; , \quad \xi_{1,2}
\in [ 0 , \, 1] \; .
\end{equation}

\subsubsection{Energy parameterisation for single-unresolved configurations}
\label{sec:SUEnergies}

\begin{figure}[b]
  \begin{center}
    \includegraphics[scale=.6]{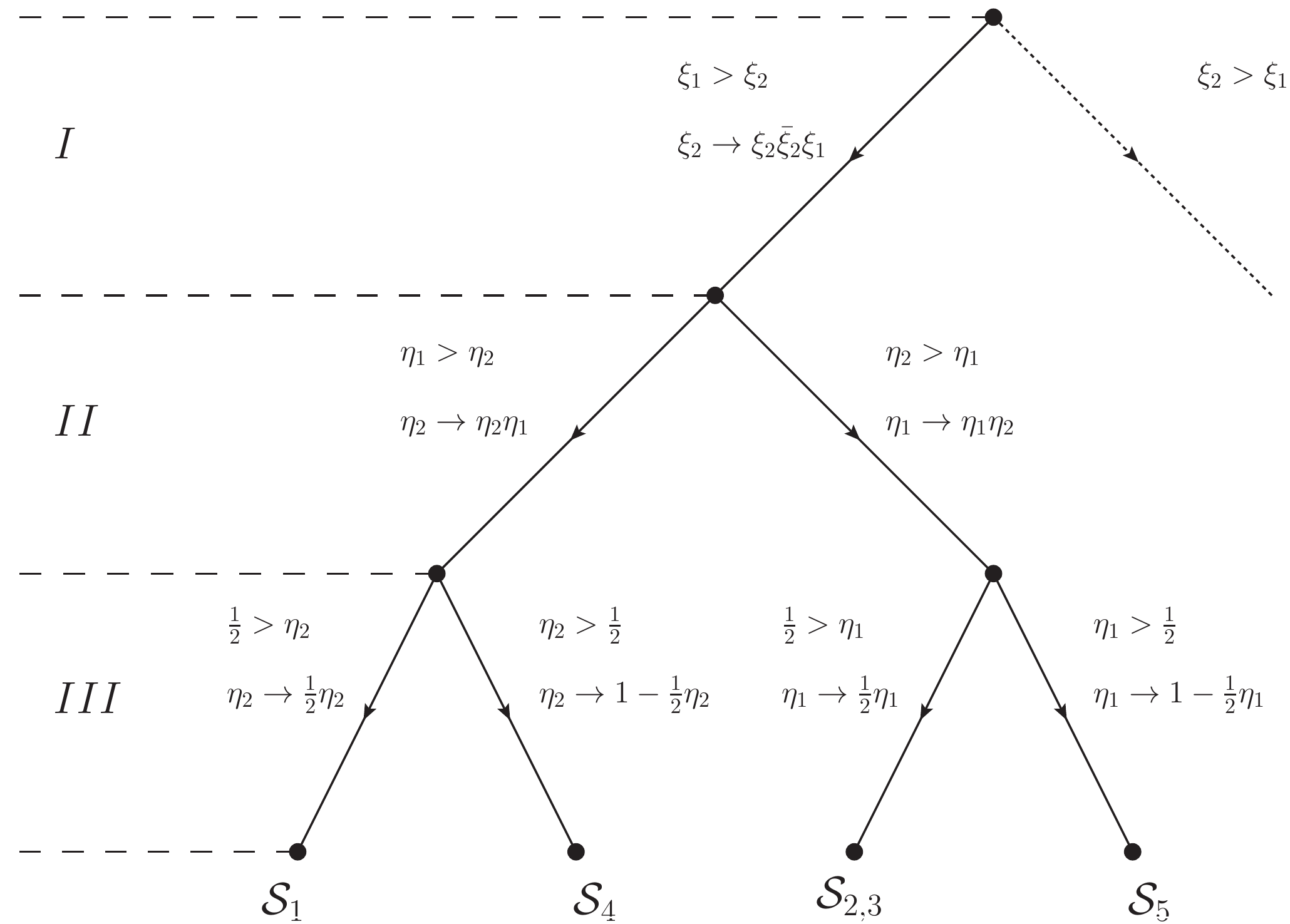}
  \end{center}
  \caption{\label{fig:decomposition} \sf Decomposition tree of the
    triple-collinear sector unresolved phase space. The
    omitted right branch of the tree corresponds to a different
    ordering of the energies of the unresolved partons, and can be
    obtained by renaming the indices of the variables, $1
    \leftrightarrow 2$. The function $\bar{\xi}_2$ is
    defined implicitly in Eq.~(\ref{eq:OneRefMomEnergies}).}
\end{figure}

The parameterisation of the angles of the unresolved momenta in the
case of a single reference momentum follows Section 4.2 of
Ref.~\cite{Czakon:2014oma}. In particular, there is
\begin{equation}
\begin{gathered}
\frac{\hat{u}_1 \cdot \hat{r}}{2} = 1-2\hat{\eta}_1 \; , \quad
\frac{\hat{u}_2 \cdot \hat{r}}{2} = 1-2\hat{\eta}_2 \; , \\[0.2cm]
\frac{\hat{u}_1 \cdot \hat{u}_2}{2} =
\frac{(\hat{\eta}_1-\hat{\eta}_2)^2}{\hat{\eta}_1+\hat{\eta}_2-2\hat{\eta}_1
  \hat{\eta}_2-2(1-2\zeta)
  \sqrt{\hat{\eta}_1(1-\hat{\eta}_1)\hat{\eta}_2(1-\hat{\eta}_2)}} \; .
\end{gathered}
\end{equation}
The phase space is further decomposed in the variables $\hat{\eta}_1$ and
$\hat{\eta}_2$ as shown in Fig.~\ref{fig:decomposition} in terms of $\eta_1$
and $\eta_2$, where we have merged sector 2 and 3 defined in
Ref.~\cite{Czakon:2014oma} as suggested in Ref.~\cite{Caola:2017dug}.
Starting at the root with $\eta_i = \hat{\eta}_i$, $\xi_i = \hat{\xi}_i$,
where $u_i^0 = \big( u_1^0 \big)_{\mathrm{max}} \hat{\xi}_i$, the first level
of decomposition corresponds to the energy parameterization in
Eq.~(\ref{eq:OneRefMomEnergies}) to factorize the soft limits.
The substitutions at level II and III factorize the collinear limits in each
sector $\mathcal{S}_i$. For example in sector $\mathcal{S}_1$ we obtain the
parameterization $\hat{\eta}_1 = \eta_1$ and $\hat{\eta}_2 = \eta_1\eta_2/2$ and
similar in the other sectors.
We consider the four sectors separately and verify the number of
subtraction terms in the single-unresolved configurations

\begin{description}

\item{\bf Sector 1} involves two independently vanishing variables
  ($\eta_2$ or $\xi_2$) in the single-unresolved kinematics, which
  corresponds to $u_2 = \alpha \, r$. The relevant resolved momenta
  are $u_1$ and $r+u_2$ ($r/z-u_2$) for a final-state (initial-state)
  reference.  $u_1$ is specified completely without any reference to
  $u_2$ thanks to the iterative energy parameterisation. The
  single-unresolved configuration is thus unique by the arguments of
  Section~\ref{sec:PhaseSpaceMapping}.

\item{\bf Sector 2,3} involves one vanishing variable for each of the
  two partons in the single-unresolved kinematics ($\eta_1$ for $u_1$,
  $\xi_2$ for $u_2$) implying that there are two single-unresolved
  configurations.

\item{\bf Sectors 4 and 5} involve two independently vanishing
  variables ($\eta_2$ for sector 4, $\eta_1$ for sector 5, or $\xi_2$
  in both sectors) in the single-unresolved kinematics, which
  corresponds to $u_2 = \alpha \, u_1$. The relevant resolved momenta
  are $r$ (if in the final state) and $u_1 + u_2$. In the iterative
  energy parameterisation, the energy of $u_1 + u_2$ depends on the
  energy of $u_2$. For this reason the single-unresolved configuration
  is not unique.

  To make the single-unresolved configuration unique, we introduce an
  alternative energy parameterisation in terms of the sum of the
  energies and their relative proportion
\begin{equation}
u_{12}^0 \equiv u_1^0 + u_2^0 \; , \quad
\xi_2 = \frac{2 u_2^0}{(u_1^0 + u_2^0)} \; .
\end{equation}
  While $\xi_2 \leq 2$, restricting its variation range to $\xi_2 \in
  [0, \, 1]$ implies $u_1^0 \geq u_2^0$. For any value of $\xi_2$, $x$
  and $z$ are monotonically decreasing functions of
  $u_{12}^0$. Let
\begin{equation}
\bar{u}_{12} = (1-\xi_2/2) \, \hat{u}_1 + \xi_2/2 \, \hat{u}_2 \; .
\end{equation}
  For a final state reference there is
\begin{equation}
\big( u_{12}^0 \big)_{\mathrm{max}} = \frac{2 \, P \cdot \tilde{r}}{P
  \cdot \bar{u}_{12} + \sqrt{\big( P \cdot \bar{u}_{12} \big)^2 -
  2 \, \bar{u}_{12}^2 \, P \cdot \tilde{r}}} \; ,
\end{equation}
  while for an initial state reference there is
\begin{equation}
\big( u_{12}^0 \big)_{\mathrm{max}} = (1 - \tilde{x}) \, \frac{2 \, p
  \cdot \tilde{r} / \tilde{x}}{(\tilde{r}/\tilde{x} + p) \cdot
  \bar{u}_{12} + \sqrt{\big( (\tilde{r}/\tilde{x} + p) \cdot
    \bar{u}_{12} \big)^2 - 2 \, \bar{u}_{12}^2 \, (1 - \tilde{x}) \,  p
    \cdot \tilde{r} / \tilde{x}}} \; .
\end{equation}
  With the energies parameterised as follows
\begin{equation}
u_1^0 = \big( u_{12}^0 \big)_{\mathrm{max}} \, \xi_1 \, (1-\xi_2/2) \; , \quad
u_2^0 = \big( u_{12}^0 \big)_{\mathrm{max}} \, \xi_1 \, \xi_2/2 \; , \quad
\xi_{1,2} \in [ 0, \, 1 ] \; ,
\end{equation}
  the integration measure is
\begin{equation}
\mathrm{d} u_1^0 \, \mathrm{d} u_2^0 = \frac{1}{2} \, \big( u_{12}^0
\big)_{\mathrm{max}}^2 \, \xi_1 \, \mathrm{d} \xi_1 \, \mathrm{d}
\xi_2 \; .
\end{equation}
  If $u_2 = \alpha \, u_1$, $\big( u_{12}^0 \big)_{\mathrm{max}}$ does
  not depend on $u_2^0$. Thus $\xi_1$ uniquely determines the resolved
  momentum $u_1 + u_2$. At the same time the rescaling of the
  reference momentum is also unique. In consequence, we have only one
  single-unresolved configuration in sector 4. There is also only one
  single-unresolved configuration at $\eta_1 = 0$ independently of
  $\xi_2$ in sector 5. However, there is a second single-unresolved
  configuration at $\xi_2 = 0$, $\eta_1 \neq 0$ as discussed below.

  The remapping of the energy variables can be introduced into the
  parameterisation of the phase space from the start, since the sectors
  4 and 5 are defined independently of the energy of the partons. In
  principle, both energy parameterisations may be used in the sector
  2,3. However, sector 1 requires the original parameterisation in order
  not to introduce additional subtraction kinematics.

\item{\bf Sector 5.} At this point, the configurations corresponding
  to $\eta_1 = 0$ and $\xi_2 = 0, \eta_1 \neq 0$ are different. This
  is due to the fact that the direction of the soft momentum $u_2$ in
  the latter case influences the direction of $u_1$ which is
  resolved. Thus, there is a second single-unresolved configuration.

\end{description}

\noindent
By the above considerations, we have one single-unresolved
configuration in sectors 1 and 4, and two single-unresolved
configurations in sectors 2,3 and 5.

\subsubsection{Two reference momenta}
\label{sec:TwoRefMom}

The following constraints allow to determine $x_{1,2}$, $z_{1,2}$ (the
classification corresponds to the position of the references $r_1$ and
$r_2$ in that order, $p$ is the second initial state momentum)
\begin{alignat*}{2}
\intertext{\bf final-final:}
&\big( P - r_1 - u_1 - r_2 - u_2 \big)^2 = \big( P - \tilde{r}_1 -
\tilde{r}_2 \big)^2 \; ,
& \quad &\big( P - r_1 - u_1 \big)^2 = \big( P - \tilde{r}_1 \big)^2 \; ,
\\[0.2cm]
&x_1 = \frac{P \cdot r_1}{(P - u_1) \cdot (r_1 + u_1)} \; ,
&&x_2 = \frac{(P - r_1/x_1) \cdot r_2}{(P - r_1 - u_1 - u_2) \cdot
  (r_2 + u_2)} \; ,
\\[0.2cm]
&x_1 = \frac{P \cdot (\tilde{r}_1 - u_1)}{(P - u_1) \cdot \tilde{r}_1} \; ,
&&x_2 = \frac{(P - \tilde{r}_1) \cdot \tilde{r}_2 - (P - x_1
  \tilde{r}_1 - u_1) \cdot u_2}{(P - x_1 \tilde{r}_1 - u_1 - u_2)
  \cdot \tilde{r}_2} \; ,
 \\[0.2cm]
&\big( u_1^0 \big)_{\mathrm{\max}} = \frac{P \cdot \tilde{r}_1}{P
  \cdot \hat{u}_1} \; ,
&&\big( u_2^0 \big)_{\mathrm{\max}} = \frac{(P - \tilde{r}_1) \cdot
  \tilde{r}_2}{(P - x_1 \tilde{r}_1 - u_1) \cdot \hat{u}_2} \; ,
\\[0.2cm]
&\mathcal{J} = \frac{x_1^{d-3} \, P \cdot \tilde{r}_1}{(P - u_1) \cdot
  \tilde{r}_1} \frac{x_2^{d-3} \, (P - \tilde{r}_1) \cdot
  \tilde{r}_2}{(P - x_1 \tilde{r}_1 - u_1 - u_2) \cdot \tilde{r}_2} \;
,
\\[0.2cm]
\intertext{\bf final-initial:}
&\big( p + r_2 - r_1 - u_1 - u_2 \big)^2 = \big( p + \tilde{r}_2 -
\tilde{r}_1 \big)^2 \; ,
& \quad &\big( p + \tilde{r}_2 - r_1 - u_1 \big)^2 = \big( p +
\tilde{r}_2 - \tilde{r}_1 \big)^2 \; ,
\\[0.2cm]
&z_2 = \frac{(p - r_1 - u_1 - u_2) \cdot (r_2 - u_2)}{(p - r_1 - u_1)
\cdot r_2} \; ,
&&x_1 = \frac{(p + z_2 r_2) \cdot r_1}{(p + z_2 r_2 - u_1) \cdot (r_1 +
  u_1)} \; ,
\\[0.2cm]
&x_1 = \frac{(p + \tilde{r}_2) \cdot (\tilde{r}_1 - u_1)}{(p +
  \tilde{r}_2 - u_1) \cdot \tilde{r}_1} \; ,
&&z_2 = \frac{(p - x_1 \tilde{r}_1 - u_1 - u_2) \cdot \tilde{r}_2}{(p -
  x_1 \tilde{r}_1 - u_1) \cdot (\tilde{r}_2 + u_2)} \; ,
\\[0.2cm]
&\big( u_1^0 \big)_{\mathrm{\max}} = \frac{(p + \tilde{r}_2) \cdot
  \tilde{r}_1}{(p + \tilde{r}_2) \cdot \hat{u}_1} \; ,
&&\big( u_2^0 \big)_{\mathrm{\max}} = (1 - \tilde{x}_2) \frac{(p - x_1
  \tilde{r}_1 - u_1) \cdot \tilde{r}_2/\tilde{x}_2}{(p +
  \tilde{r}_2/\tilde{x}_2 - x_1 \tilde{r}_1 - u_1) \cdot \hat{u}_2} \; ,
\\[0.2cm]
&\mathcal{J} = \frac{x_1^{d-3} \, (p + \tilde{r}_2) \cdot \tilde{r}_1}{(p
+ \tilde{r}_2 - u_1) \cdot \tilde{r}_1} \frac{(p - x_1 \tilde{r}_1 -
u_1) \cdot \tilde{r}_2}{(p - x_1 \tilde{r}_1 - u_1 - u_2) \cdot
\tilde{r}_2} \; ,
\\[0.2cm]
\intertext{\bf initial-final:}
&\big( r_1 + p - u_1 - r_2 - u_2 \big)^2 = \big( \tilde{r}_1 + p -
\tilde{r}_2 \big)^2 \; ,
& \quad &\big( r_1 + p - u_1 \big)^2 = \big( \tilde{r}_1 + p \big)^2
\; ,
\\[0.2cm]
&z_1 = \frac{(r_1 + p) \cdot (r_1 - u_1)}{p \cdot r_1} \; ,
&&x_2 = \frac{(z_1 r_1 + p) \cdot r_2}{(r_1 + p - u_1 - u_2) \cdot
  (r_2 + u_2)} \; ,
\\[0.2cm]
&z_1 = \frac{(p - u_1) \cdot \tilde{r}_1}{p \cdot (\tilde{r}_1 + u_1)}
\; ,
&&x_2 = \frac{(\tilde{r}_1 + p) \cdot \tilde{r}_2 - (\tilde{r}_1/z_1 +
  p - u_1) \cdot u_2}{(\tilde{r}_1/z_1 + p - u_1 - u_2) \cdot
  \tilde{r}_2} \; ,
\\[0.2cm]
&\big( u_1^0 \big)_{\mathrm{\max}} = (1 - \tilde{x}_1) \frac{p \cdot
  \tilde{r}_1/\tilde{x}_1}{(\tilde{r}_1/\tilde{x}_1 + p) \cdot
  \hat{u}_1} \; ,
&&\big( u_2^0 \big)_{\mathrm{\max}} = \frac{(\tilde{r}_1 + p) \cdot
  \tilde{r}_2}{(\tilde{r}_1/z_1 + p - u_1) \cdot \hat{u}_2} \; ,
\\[0.2cm]
&\mathcal{J} = \frac{p \cdot \tilde{r}_1}{(p - u_1) \cdot \tilde{r}_1}
\frac{x_2^{d-3} \, (\tilde{r}_1 + p) \cdot
  \tilde{r}_2}{(\tilde{r}_1/z_1 + p - u_1 - u_2) \cdot \tilde{r}_2} \;
,
\\[0.2cm]
\intertext{\bf initial-initial:}
&\big( r_1 + r_2 - u_1 - u_2 \big)^2 = \big( \tilde{r}_1 + \tilde{r}_2
\big)^2 \; ,
& \quad &\big( r_1 + \tilde{r}_2 - u_1 \big)^2 = \big( \tilde{r}_1 +
\tilde{r}_2 \big)^2 \; ,
\\[0.2cm]
&z_2 = \frac{(r_1 - u_1 - u_2) \cdot (r_2 - u_2)}{(r_1 - u_1) \cdot
  r_2} \; ,
&& z_1 = \frac{(z_2 r_2 - u_1) \cdot (r_1 - u_1)}{z_2 r_2 \cdot r_1}
\; ,
\\[0.2cm]
&z_1 = \frac{(\tilde{r}_2 - u_1) \cdot \tilde{r}_1}{\tilde{r}_2 \cdot
  (\tilde{r}_1 + u_1)} \; ,
&& z_2 = \frac{(\tilde{r}_1/z_1 - u_1 - u_2) \cdot
  \tilde{r}_2}{(\tilde{r}_1/z_1 - u_1) \cdot (\tilde{r}_2 + u_2)} \; ,
\\[0.2cm]
&\big( u_1^0 \big)_{\mathrm{\max}} = (1 - \tilde{x}_1)
\frac{\tilde{r}_2 \cdot
  \tilde{r}_1/\tilde{x}_1}{(\tilde{r}_1/\tilde{x}_1 + \tilde{r}_2)
  \cdot \hat{u}_1} \; ,
&& \big( u_2^0 \big)_{\mathrm{\max}} = (1 - \tilde{x}_2)
\frac{(\tilde{r}_1/z_1 - u_1) \cdot
  \tilde{r}_2/\tilde{x}_2}{(\tilde{r}_1/z_1 + \tilde{r}_2/\tilde{x}_2
  - u_1) \cdot \hat{u}_2} \; ,
\\[0.2cm]
&\mathcal{J} = \frac{\tilde{r}_2 \cdot \tilde{r}_1}{(\tilde{r}_2 -
  u_1) \cdot \tilde{r}_1} \frac{(\tilde{r}_1/z_1 - u_1) \cdot
  \tilde{r}_2}{(\tilde{r}_1/z_1 - u_1 - u_2) \cdot \tilde{r}_2} \; .
\end{alignat*}
They have the following properties
\begin{enumerate}

\item the Born configuration is well defined for any full
  configuration, i.e. $x_{1,2}, z_{1,2}, \tilde{q}^0 \geq 0$, and $x_1
  \leq 1$, $z_{1,2} \leq 1$,

\item $x_1$, $z_1$ are monotonically decreasing functions of $u_1^0$
  independent of $u_2$,

\item $x_2$, $z_2$ are monotonically decreasing functions of $u_2^0$
  at fixed $u_1$.

\end{enumerate}
In consequence, an iterative energy parameterisation with $u_1^0$
determined in the range $\big[ 0, \, \big( u_1^0
\big)_{\mathrm{max}} \big]$, followed by $u_2^0$ in the range $\big[
0, \, \big( u_2^0 \big)_{\mathrm{max}}(u_1) \big]$ covers the full
phase space. The maxima of the energies, $\big( u_i^0
\big)_{\mathrm{max}}$, are obtained at $x_i = 0$ or $z_i =
\tilde{x}_i$.

Imposing an ordering of the energies, $u_1^0 \geq u_2^0$, for a given
phase space parameterisation is compensated by adding the contribution
of the phase space for the parameterisation corresponding to swapped
references. This covers the full phase space if and only if the condition
$u_1^0 \geq u_2^0$ is applied in the same frame in both
cases. Unfortunately, if at least one of the references is in the
initial state, the chosen parameterisations lead to different Born
frames upon swapping the references. Therefore, the ordering of the
energies cannot be imposed in the Born frame. Instead, we apply the
parameterisations in a fixed frame with respect to the lab. The full
phase space is then correctly covered with the energy parameterisation
\begin{equation}
u_1^0 = \big( u_1^0 \big)_{\mathrm{max}} \, \xi_1 \; , \quad
u_2^0 = \big( u_1^0 \big)_{\mathrm{max}} \, \xi_1 \xi_2 \min \Bigg[
1, \, \frac{1}{\xi_1} \frac{\big( u_2^0 \big)_{\mathrm{max}}}{\big(
u_1^0 \big)_{\mathrm{max}}} \Bigg] \; , \quad
\xi_{1,2} \in [ 0, \, 1 ] \; .
\end{equation}
The parameterisation of the angles of the unresolved momenta follows
Section 4.3 of Ref.~\cite{Czakon:2014oma}.


\subsection{Reduction to four dimensions}

The construction of local subtraction terms following the strategy of
sector decomposition (see Ref.~\cite{Czakon:2014oma} for details)
yields integrable expressions in CDR. The different cross section
contributions are Laurent-series expansions with poles in $\ep$ (for
the CDR parameter $d = 4-2\ep$) whose sum is finite due to the
finiteness of the (next-to-)next-to-leading order cross section. The
construction in $d$ dimensions is straightforward but computationally
cumbersome due to: 1) the necessity of including higher order
$\ep$-expansion terms of matrix elements; 2) the growth with
multiplicity of the number of effective dimensions of the phase space
parameterisation. In Ref.~\cite{Czakon:2014oma}, it has been shown
that a four-dimensional formulation of the subtraction scheme can be
given by introducing additional corrections such that the
single-unresolved ($SU$) and double-unresolved ($DU$) contributions to
the next-to-next-to-leading order cross section are separately
finite. In order to determine these corrections, separately finite
sets of contributions must be identified. For the sake of completeness
we first review the necessary notation.

The hadronic cross section is given by the collinear factorisation
expression
\begin{multline}
\sigma_{h_1h_2}(P_1, P_2) = \\[0.2cm] \sum_{ab} \iint_0^1
\mathrm{d}x_1 \mathrm{d}x_2 \, f_{a/h_1}(x_1, \mF) \,
f_{b/h_2}(x_2, \mF) \, \hat{\sigma}_{ab}(x_1P_1, x_2P_2;
\, \as(\mR), \, \mR, \, \mF) \; ,
\end{multline}
where $P_{1,2}$ are the momenta of the hadrons $h_{1,2}$, while
$p_{1,2} = x_{1,2} P_{1,2}$ are the momenta of the
partons. $f_{a/h}(x,\mF)$ is the PDF of parton $a$ within the hadron
$h$, at the factorisation scale $\mu_F$. The partonic cross section
can be systematically expanded in the strong coupling constant
$\alpha_s$
\begin{align}
  \sigh_{ab} = \sigh_{ab}^{(0)}+\sigh_{ab}^{(1)}+\sigh_{ab}^{(2)} +
  \dots\; .
\end{align}
The cross sections $\sigh_{ab}^{(i)}$ are sums of several
contributions differing by the final state multiplicity, parton
flavours and the number of loops of the involved matrix elements. For
instance
\begin{align}
  \sigh_{ab}^{(1)} &= \sigh_{ab}^R+\sigh_{ab}^V+\sigh_{ab}^C\, ,\\[.2cm]
  \sigh_{ab}^{(2)} &= \sigh_{ab}^{RR}+\sigh_{ab}^{RV}+\sigh_{ab}^{VV}
                     +\sigh_{ab}^{C1}+\sigh_{ab}^{C2}\; .
\end{align}
Here, the superscript ``$R$'' denotes emission of an additional parton
w.r.t.\ to the Born final state, ``$V$'' denotes a virtual-loop
integration, while ``$C$'' a convolution with Altarelli-Parisi
splitting kernels. Precise definitions are given in
Appendix~\ref{app:notation}.

After introduction of sectors followed by the derivation of the
subtraction and integrated subtraction terms, the next-to-leading
order real-emission contribution is decomposed as follows
\be
  \sigh^R = \sigh^R_F + \sigh^R_U \; ,
\ee
where $\sigh^R_F$ contains the $(n+1)$-particle tree-level matrix
elements together with appropriate subtraction terms, while
$\sigh^R_U$ contains the respective integrated subtraction terms and
$n$-particle tree-level matrix elements.

In general, infrared divergences can be factorised from virtual
amplitudes as follows
\be
\label{eq:FinRem}
  |\cm_n \ra = \bm{\mathrm{Z}}(\ep,\{p_i\},\{m_i\},\mu_R) \, |\cf_n \ra \; ,
\ee
where the infrared renormalisation constant $\bm{\mathrm{Z}}$ is an
operator in color space, and depends on the momenta $\{p_i\}=\{p_1,
..., p_n\}$ and masses $\{m_i\}=\{m_1,...,m_n\}$ of the external
partons. The finite remainder, $|\cf_n \ra$, has a well-defined limit
when $\ep \rightarrow 0$. Expanding equation (\ref{eq:FinRem}) in a
series in $\as$ yields
\begin{align}
 |\mathcal{M}_n^{(0)}\ra
&=|\mathcal{F}_n^{(0)}\ra \; , \label{eq:ExpandZM0} \\[0.2cm]
|\mathcal{M}_n^{(1)}\ra &=
\bm{\mathrm{Z}}^{(1)}|\mathcal{M}_n^{(0)}\ra
+|\mathcal{F}_n^{(1)}\ra \; , \label{eq:ExpandZM1}\\[0.2cm]
|\mathcal{M}_n^{\left(2\right)}\ra &=
\bm{\mathrm{Z}}^{\left(2\right)}|\mathcal{M}_n^{(0)}\ra +
\bm{\mathrm{Z}}^{(1)}|\mathcal{F}_n^{(1)}\ra +
|\mathcal{F}_n^{\left(2\right)}\ra \nn\\[0.2cm]
&=\left(
\bm{\mathrm{Z}}^{\left(2\right)}-\bm{\mathrm{Z}}^{(1)}
\bm{\mathrm{Z}}^{(1)}\right)|\mathcal{M}_n^{(0)}\ra
+ \bm{\mathrm{Z}}^{(1)}|\mathcal{M}_n^{(1)}\ra +
|\mathcal{F}_n^{\left(2\right)}\ra \; ,
\label{eq:ExpandZM2}
\end{align}
with $\bm{\mathrm{Z}} = \bm{1} + \bm{\mathrm{Z}}^{(1)} +
\bm{\mathrm{Z}}^{(2)} + \mathcal{O}(\alpha_s^3)$. This decomposition
translates into a decomposition of the virtual contribution at
next-to-leading order
\be
  \sigh^V = \sigh^V_F + \sigh^V_U \; ,
\ee
where $\sigh^V_F$ contains the $n$-particle one-loop finite remainders,
while $\sigh^V_U$ contains the $n$-particle tree-level matrix elements
of the $\bm{\mathrm{Z}}^{(1)}$ operator.

In consequence, there are three separately finite contributions in
the next-to-leading order case
\begin{align}
  \sigh^R_F\;,\quad\sigh^V_F\;,\quad\sigh_U =\sigh^R_U+\sigh^V_U + \sigh^C \;.
\end{align}
In each separately finite contribution, it is allowed to take the
$\ep \rightarrow 0$ limit by removing higher-order terms in the
$\ep$-expansion of the matrix elements and reducing the dimension of
the resolved momenta to four. This is the essence of the
four-dimensional formulation of the subtraction scheme.

This construction can be extended to next-to-next-to-leading order, which
yields the following decompositions
\begin{align}
  \sigh^{RR} &= \sigh^{RR}_F + \sigh^{RR}_{SU} + \sigh^{RR}_{DU}\; ,
  \\[0.2cm]
  \sigh^{RV} &=
  \sigh^{RV}_F+\sigh^{RV}_{SU}+\sigh^{RV}_{FR}+\sigh^{RV}_{DU}\; ,
  \\[0.2cm]
  \sigh^{VV} &= \sigh^{VV}_F + \sigh^{VV}_{FR} + \sigh^{VV}_{DU} \; ,
  \\[0.2cm]
  \sigh^{C1} &= \sigh^{C1}_{SU} + \sigh^{C1}_{DU} \; , \\[0.2cm]
  \sigh^{C2} &= \sigh^{C2}_{FR} + \sigh^{C2}_{DU} \; .
\end{align}
The different contributions are identified as
follows. $\sigh^{RR,RV,VV}_F$ contain the same multiplicity and
number-of-loops finite-remainder matrix elements as
$\sigh^{RR,RV,VV}$ together with appropriate subtraction terms to
make them integrable (unnecessary for $\sigh^{VV}$). The subscript
``$FR$'' (Finite Remainder) denotes the remaining contributions
containing at most $n$-particle one-loop finite-remainder matrix
elements. The subscript ``$SU$'' (Single Unresolved) denotes the
remaining contributions containing at most $(n+1)$-particle tree-level
matrix elements together with appropriate subtraction terms to make
them integrable. Finally, the subscript ``$DU$'' (Double Unresolved)
denotes the remaining contributions containing only $n$-particle
tree-level matrix elements. The $FR$, $SU$ and $DU$ contributions
contain explicit poles in $\ep$ due to
$\bm{\mathrm{Z}}^{(1,2)}$-operator insertions and integration over
subtraction terms of the $F$-contributions (and $SU$-contributions in
the case of $DU$-contributions).

By construction, three contributions are separately finite
\begin{align}
  \sigh^{RR}_F\;,\quad \sigh^{RV}_F\;,\quad \sigh^{VV}_F\;.
\end{align}
The finiteness of the next-to-next-to-leading order cross section
implies that
\begin{align}
  \sigh_{DU} + \sigh_{SU} + \sigh_{FR} = \text{finite} \; ,
\end{align}
where
\be
\begin{aligned}
 \sigh_{FR} &= \sigh^{RV}_{FR}+\sigh^{VV}_{FR}+\sigh^{C2}_{FR}\;,\\[0.2cm]
 \sigh_{SU} &= \sigh^{RR}_{SU}+\sigh^{RV}_{SU}+\sigh^{C1}_{SU}\;,\\[0.2cm]
 \sigh_{DU} &= \sigh^{RR}_{DU}+\sigh^{RV}_{DU}+\sigh^{C1}_{DU}+\sigh^{VV}_{DU}
              +\sigh^{C2}_{DU}\;.
\end{aligned}
\ee
Following the argument of Ref.~\cite{Czakon:2014oma}, $\sigh_{FR}$ is
separately finite due to the finiteness of the next-to-leading order
cross section. Indeed, $\sigh_{FR}$ is given by the same expressions
as $\sigh_{U}$ once tree-level amplitudes are replaced by one-loop
finite remainders in the latter. This leaves $\sigh_{DU} + \sigh_{SU}$
to be finite.

Unfortunately, it turns out that $\sigh_{DU}$ and $\sigh_{SU}$ are both
separately divergent despite having different multiplicity resolved
final states. A four-dimensional formulation of the subtraction scheme
is only obtained under the assumption that a $\sigh_{HV}$, the 't
Hooft-Veltman corrections contribution, linear in the infrared safe
measurement function $F_n$ exists such that
\begin{align}
  \sigt_{SU} &= \sigh_{SU} - \sigh_{HV} \; , \\[0.2cm]
  \sigt_{DU} &= \sigh_{DU} + \sigh_{HV} \; ,
\end{align}
are separately finite. An appropriate $\sigh_{HV}$ has been
constructed in Ref.~\cite{Czakon:2014oma}. Here, we present a
different construction which exploits the fact that $\sigh_{SU}$ is
finite for a next-to-leading order measurement function, i.e.\ for
$F_n = 0$. The approach relies on the idea that, since the cut on the
additional radiation in $F_{n+1}$ is arbitrary, the divergences in
$\sigh_{SU}$ with a next-to-next-to-leading order measurement
function, i.e.\ for $F_n \neq 0$, can, in fact, be described with
$n$-particle kinematics and matrix elements. It should thus be
possible to systematically identify them and subsequently shift them
to $\sigh_{DU}$.

\subsubsection{'t Hooft-Veltman corrections}
\label{sec:tHVCor}

The replacement of the next-to-next-to-leading order measurement
function with a next-to-leading order measurement function, i.e.\ one
with $F_{n+1} \neq 0$ and $F_n = 0$, turns $\sigh_{SU}$ into the $\sigh_U$
of an $(n+1)$-particle next-to-leading order cross section with
\begin{align}
  \sigh^{RU}_{SU} \; \to \; \sigh^R_U\;,\quad \sigh^{RV}_{SU} \; \to
  \; \sigh^V_U\;,\quad \sigh^{C1}_{SU} \; \to \; \sigh^C\,.
\end{align}
At next-to-leading order, the contributions $\sigh^R_U$, $\sigh^V_U$
and $\sigh^C$ may be written in the following form
\begin{align}
  \sigh^R_U &= \int \dd \Phi^{(d)}_{n+1} \; \mathcal{I}^R_{n+1}
              F_{n+1} \; , \\[0.2cm]
  \sigh^V_U &= \int \dd \Phi^{(d)}_{n+1} \; \mathcal{I}^V_{n+1}
              F_{n+1} \; , \\[0.2cm]
  \sigh^C   &= \int \dd \Phi^{(d)}_{n+1} \; \mathcal{I}^C_{n+1}
              F_{n+1} \; .
\end{align}
Notice that $\mathcal{I}^R_{n+1}$ is given by a sum of contributions
of individual sectors, which makes the factorisation of the
$d$-dimensional phase space measure $\dd \Phi^{(d)}_{n+1}$
non-trivial.  With the phase space parameterisations of
Section~\ref{sec:PhaseSpace}, this factorisation can, nevertheless, be
achieved explicitly in each sector.

Let us now consider the structure of the $\ep$-expansions of the
integrands $\mathcal{I}^c_{n+1}$, $c \in \{R,V,C\}$, while keeping the
exact $\ep$-dependence of the occurring matrix elements. We have
\begin{align}
  \mathcal{I}^R_{n+1} &= \frac{\mathcal{I}^{R (-2)}_{n+1}}{\ep^2} +
                        \frac{\mathcal{I}^{R (-1)}_{n+1}}{\ep} +
                        \mathcal{I}^{R (0)}_{n+1} + \order{\ep} \; ,
  \\[0.2cm]
  \mathcal{I}^V_{n+1} &= \frac{\mathcal{I}^{V (-2)}_{n+1}}{\ep^2} +
                        \frac{\mathcal{I}^{V (-1)}_{n+1}}{\ep} \; ,
  \\[0.2cm]
  \mathcal{I}^C_{n+1} &= \frac{\mathcal{I}^{C (-1)}_{n+1}}{\ep} +
                        \mathcal{I}^{C (0)}_{n+1}  + \order{\ep} \; ,
\end{align}
where each $\mathcal{I}^{c(i)}_{n+1}$ is a function of $\ep$ through
the $\ep$-dependence of the tree-level matrix elements only.  The
simple structure of $\mathcal{I}^V_{n+1}$ is due to the fact that it
is given by matrix elements of the $\mathbf{Z}^{(1)}$-operator which
we chose to be defined in the \msbar scheme, where it consists of pure
poles. Even though collinear factorisation is also performed in the
\msbar scheme, $\mathcal{I}^C_{n+1}$ does have a non-trivial
$\ep$-expansion for $\mu_R \neq \mu_F$ because of the expansion of the
pre-factor $(\mu_R^2/\mu_F^2)^\ep$ in \eqref{eq:NLOXS}. The finiteness
of the next-to-leading order cross section implies
\be
  \sum_c \int \dd \Phi^{(d)}_{n+1} \left[
    \frac{\mathcal{I}^{c (-2)}_{n+1}}{\ep^2} +
    \frac{\mathcal{I}^{c (-1)}_{n+1}}{\ep} \right]F_{n+1} \equiv
  \sum_c \mathcal{I}^c = 0 \; .
  \label{eq:finiteNLO}
\ee
This analysis translates directly to $\sigh_{SU}$ with an appropriate
change of superscripts.

\paragraph{Parameterised measurement function.}

Let us introduce a family of measurement functions $F^\alpha_m$, $m
\in \{n,n+1,n+2\}$ with the following properties
\begin{itemize}
  \item $F^\alpha_m$ is infrared safe;
  \item $F^{\alpha \neq 0}_n = 0$ and $F^{\alpha = 0}_n \neq 0$.
\end{itemize}
Hence, $\alpha \neq 0$ corresponds to a next-to-leading order
calculation within a next-to-next-to-leading order setup, while
$\alpha = 0$ corresponds to the general next-to-next-to-leading order
calculation. Since we only consider single- and double-unresolved
contributions, it is not necessary to define $F^\alpha_{n+2}$. In
order to identify the 't Hooft-Veltman corrections, a particularly
useful realisation is given by
\be
  F^\alpha_{n+1} = F_{n+1} \theta(\alpha_\eta-\alpha)
  \theta(\alpha_\xi-\alpha) \equiv F_{n+1} \theta_{\eta}\theta_{\xi}
  \equiv F_{n+1} \theta_\alpha \; ,
\ee
with $F^\alpha_n$ obtained from $F^\alpha_{n+1}$ by taking soft
and/or collinear limits. $\alpha_\eta$ and $\alpha_\xi$ are a set of
global infrared sensitive variables
\begin{align}
  &\alpha_\eta = \min_{ij} \eta_{ij} \; ,
  &\text{with} \quad
  &\eta_{ij} = \frac{1}{2}\left(1 -\cos \theta_{ij}\right) \; , \\[0.2cm]
  &\alpha_\xi = \min_i \xi_i \; ,
  &\text{with} \quad
  &\xi_{i} = \frac{p_i^0}{E_\text{norm}} \; ,
\end{align}
where $\theta_{ij}$ is the angle between two parton momenta, and
$p_i^0$ is a parton energy.  The variable $\alpha_\eta$ measures the
minimal angle between any two partons $i$ and $j$, while $\alpha_\xi$
measures the minimal energy of the partons with respect to some
arbitrary energy scale $E_\text{norm}$. For $\alpha \neq 0$,
$F^\alpha_m$ is a well-defined next-to-leading order measurement
function.  For $\alpha = 0$, it corresponds to the original
next-to-next-to-leading order measurement function.

\paragraph{Identification of 't Hooft-Veltman corrections.}

Using the parameterised measurement function, the
next-to-next-to-leading order version of Eq.~\eqref{eq:finiteNLO}
takes the form
\be
  \sum_c \int \dd \Phi^{(d)}_{n+1} \; \left[
    \frac{\mathcal{I}^{c (-2)}_{n+1}}{\ep^2} +
    \frac{\mathcal{I}^{c (-1)}_{n+1}}{\ep} \right] F^{\alpha \neq
    0}_{n+1} \equiv \sum_c \mathcal{I}^c = 0 \; ,
\ee
with $c \in \{RR,RV,C1\}$. Considering the full
next-to-next-to-leading order case, we can schematically
write the contributions $\sigh^c_{SU}$ in the following form
\begin{align}
  \sigh^c_{SU}&= \int \dd \Phi^{(d)}_{n+1} \left[\mathcal{I}^c_{n+1} F_{n+1} +
                \mathcal{I}^c_{n} F_{n} \right] \\[0.2cm]
            &= \int \dd \Phi^{(d)}_{n+1} \left\{
               \left[\frac{\mathcal{I}^{c (-2)}_{n+1}}{\ep^2} +
                        \frac{\mathcal{I}^{c (-1)}_{n+1}}{\ep} +
                        \mathcal{I}^{c (0)}_{n+1} \right]F_{n+1}+
               \left[\frac{\mathcal{I}^{c (-2)}_{n}}{\ep^2} +
                        \frac{\mathcal{I}^{c (-1)}_{n}}{\ep} +
                        \mathcal{I}^{c (0)}_{n} \right]F_{n}
               \right\} \nn\\&\quad+ \order{\ep} \; .
\end{align}
The integrands $\mathcal{I}^c_n = \sum\limits_{i=-2}^\infty
\mathcal{I}^{c(i)}_n $ represent the subtraction terms that regulate
the $n$-particle limit of $\mathcal{I}^c_{n+1}$. Here, following the
discussion of the next-to-leading order case,
$\mathcal{I}^{c(i)}_{n+1}$ contain the unexpanded $(n+1)$-particle
matrix elements. Hence, $\mathcal{I}^{c(i)}_{n}$ consist of the
appropriate unexpanded factorisation formulae for $(n+1)$-particle
matrix elements in the single-soft and collinear limits.

Consider now the difference $\sigh^c_{SU} - \mathcal{I}^c$. By
reshuffling terms and neglecting $\mathcal{O}(\ep)$ contributions, it
can be written as
\begin{align}
\sigh^c_{SU} - \mathcal{I}^c &=
  \int \dd \Phi^{(d)}_{n+1} \left[
   \frac{\mathcal{I}^{c(-2)}_{n+1}F_{n+1}+ \mathcal{I}^{c(-2)}_{n}F_{n}}{\ep^2}
  +\frac{\mathcal{I}^{c(-1)}_{n+1}F_{n+1}+ \mathcal{I}^{c(-1)}_{n}F_{n}}{\ep}
   \right]\left(1-\theta_\alpha\right)\nn\\[0.2cm]
  &+\int \dd \Phi^{(d)}_{n+1}\left[\mathcal{I}^{c(0)}_{n+1}F_{n+1}+
    \mathcal{I}^{c(0)}_{n}F_{n}\right]
   +\int \dd \Phi^{(d)}_{n+1}\left[\frac{\mathcal{I}^{c(-2)}_{n}}{\ep^2}+
    \frac{\mathcal{I}^{c(-1)}_{n}}{\ep}\right]F_n
    \theta_\alpha\nn\\[0.2cm] &\equiv Z^c(\alpha) + C^c +
                                        N^c(\alpha) \; .
\end{align}
The integral $C^c$ neither depends on the parameter $\alpha$ nor
contains poles in $\ep$.  The integrand of $Z^c(\alpha)$ is integrable
while the phase space volume is restricted by $1-\theta_\alpha$. The
phase space volume thus vanishes in the $\alpha\to 0$ limit and so
does $Z^c(\alpha)$. Finally, the phase space integral in
\be
  N^c(\alpha) = \int \dd
  \Phi^{(d)}_{n+1}\left[\frac{\mathcal{I}^{c(-2)}_{n}}{\ep^2} +
  \frac{\mathcal{I}^{c(-1)}_{n}}{\ep}\right]F_n
  \theta_\alpha \; ,
\ee
contains integrations over the angle and energy variables which might
give rise to singularities regulated by $\alpha$. In particular, after
sector decomposition, the only singularities in a given phase space
sector are due to soft and collinear limits of the unresolved
parton momenta. In consequence, we can safely take the limit $\alpha
\to 0$, if neither $\alpha_\eta$ nor $\alpha_\xi$ correspond to
unresolved partons. Hence, the general contribution to $N^c(\alpha)$
contains terms regular at $\alpha \to 0$ and integrals of the form
\be
\begin{split}
  \int_0^1 \frac{\dd x }{x^{1+a\ep}} & \theta(x - f(x)\alpha) \\[0.2cm]
&= \int_0^1 \frac{\dd x }{x^{1+a\ep}} \theta(x - f(0)\alpha) +
  \int_0^1 \frac{\dd x }{x^{1+a\ep}} \left(\theta(x -
  f(x)\alpha)-\theta(x-f(0)\alpha)\right) \\[0.2cm]
&= \int_0^1 \frac{\dd x }{x^{1+a\ep}} \theta(x - f(0)\alpha)
  + \order{\alpha} \\[0.2cm]
&= -\frac{1-(f(0) \alpha)^{-a\ep}}{a\ep} + \order{\alpha} \; ,
\end{split}
\ee
where $x$ is one of $\eta_1$, $\eta_2$, $\xi_1$, $\xi_2$. Expansion in
$\ep$ yields the power-log series
\be
  N^c(\alpha) = \sum_{k=0}^{l_\text{max}} \ln^k(\alpha) N^c_k(\alpha)
  \; , \label{ch:st:eq:NCalpha}
\ee
where $N^c_k(\alpha)$ are regular at $\alpha \to 0$.

The modified $SU$ contributions are used as follows
\be
  \sigh_{SU} = \sigh_{SU} - \underbrace{\sum_c \mathcal{I}^c}_{ = 0 }
             = \sum_c \left(\sigh^c_{SU} - \mathcal{I}^c\right)
             = \sum_c \left( Z^c(\alpha) + C^c + N^c(\alpha)\right) \;.
\ee
The left-hand side is independent of $\alpha$ and, therefore, the
right-hand side has to be independent as well. Since $Z^c(\alpha)$ are
regular functions of $\alpha$ which vanish in the limit $\alpha \to
0$, the logarithms appearing in \eqref{ch:st:eq:NCalpha} have to
cancel across the different contributions $c$.  In the limit $\alpha
\to 0 $, we thus find that the poles that do not cancel within
$\sigh_{SU}$
are given by
\be
  \sum_c N^c_0(0) \equiv \sigh_{HV} \;.
\ee
Thus, subtracting $\sigh_{HV}$ from $\sigh_{SU}$ yields a finite quantity where
all poles cancel.  Since all terms in $\sigh_{HV}$ are proportional to
$F_n$, $\sigh_{HV}$ can be added to $\sigh_{DU}$.  By the finiteness
of the next-to-next-to-leading order cross section it follows that
\be
  \sigt_{SU} = \sigh_{SU} - \sigh_{HV} \; , \quad \text{and} \quad
  \sigt_{DU} = \sigh_{DU} + \sigh_{HV} \; ,
\ee
are separately finite.

The formal manipulations of the different contributions must be performed in
$d$ dimensions. However, after this procedure, the 't Hooft-Veltman
regularisation discussed in Section 8 of Ref.~\cite{Czakon:2014oma} can be
applied yielding the desired four-dimensional formulation of the subtraction
scheme. A collection of the required 't Hooft-Veltman corrections can
be found in Appendix~\ref{app:tHVcor}.


\subsection{Implementation and tests}
\label{sec:implementation}

We have implemented the complete subtraction scheme in the C++ library
\textsc{Stripper}. In principle, the library provides sufficient
functionality to evaluate NNLO QCD corrections to any process in the
Standard Model. In practice, it requires appropriate matrix elements
at tree-level (including up to double correlations in color and/or
spin), one-loop level (including single correlations in color or
spin), and two-loop level. Tree-level matrix elements for arbitrary
Standard Model processes are provided by default by the
\textsc{Fortran} library \cite{avhlib} introduced in
Ref.~\cite{Bury:2015dla}. The code generates amplitudes on-the-fly for
arbitrary polarisations (helicities) and color configurations of
external states and evaluates them numerically in double precision. We
note that, while completely general, this is slower than dedicated
analytic expressions for parton-scattering processes at low
multiplicity. The five-point one-loop matrix element values that were
required for our computation of jet rates have been obtained using the
\textsc{NJet} C++ library presented in Refs.~\cite{Badger:2010nx,
Badger:2012pg}. However, the general numerical-unitarity algorithm
implemented in the library turned out to be too slow and unstable for
our purposes. Instead, we have used the analytic formulae
\cite{Bern:1993mq, Bern:1994fz, Kunszt:1994nq} for five-parton matrix
elements implemented in \textsc{NJet}. The two-loop amplitudes have
been taken from Ref.~\cite{Broggio:2014hoa}, which is based on
Refs.~\cite{Glover:2003cm, Glover:2004si, Bern:2003ck,
DeFreitas:2004kmi, Bern:2002tk}.

The correctness of the results obtained with \textsc{Stripper} depends
on the correctness of the matrix elements, splitting and soft
functions, $d$-dimensional and four-dimensional phase spaces, and,
finally, the 't Hooft-Veltman corrections. Apart from matrix elements,
the majority of these contributions is involved in the evaluation of
the NNLO QCD corrections to hadronic top-quark pair production. With
the new implementation, we were able to reproduce the results of
Refs.~\cite{Czakon:2015owf, Czakon:2016dgf}, which have recently been
confirmed in Ref.~\cite{Catani:2019hip}. On the other hand, our most
recent results \cite{Behring:2019iiv} for this process including
Narrow-Width-Approximation top-quark decays in the di-lepton channel
have only been obtained with the new version of \textsc{Stripper}.

A final test of the phase space implementation and the 't
Hooft-Veltman corrections has been performed by calculating the
single-jet inclusive double-differential ($p_T$, $|y|$) cross section
in the pure-gluon channel for proton-proton collisions at 7 TeV
center-of-mass energy, jets defined with the anti-$k_T$ algorithm with
$R = 0.7$ and with the scale $\mu_R = \mu_F = \mu = p_{T,1}/2,
p_{T,1}, 2p_{T,1}$ for the MMHT2014nnlo68cl PDF set. In this case, we
have found perfect agreement within the statistical errors estimated
at below 1\% with results obtained with \textsc{NNLOjet} (private
communication). This test covers all aspects of our software necessary
for the evaluation of the cross sections in the remaining channels
involving quarks.


\section{Single-jet inclusive rates for LHC @ 13 TeV}
\label{sec:results}

In this section, we present our results for the single-jet inclusive
differential distributions in the jet transverse momentum ($p_T$) in
several jet rapidity ($|y|$) slices. We assume an initial state
corresponding to proton-proton collisions at 13 TeV center-of-mass
energy and use the central PDF4LHC15\_nnlo PDF set to obtain the
parton densities. The strong coupling constant running corresponds to
this PDF set as well. Jets are identified with the anti-$k_T$ jet
algorithm with $R = 0.7$, $p_T > 114$ GeV and $|y| < 4.7$. Every jet
identified in a given final-state parton configuration (event) is
input into the appropriate rapidity-slice histogram with a weight that
corresponds to a cross section contribution evaluated with the scale
$\mu_R = \mu_F = \mu \in \{ p_T,2p_T,4p_T \}$. The weight
corresponding to $\mu = 2p_T$ is taken as the central value of the
prediction, while the remaining values are used to estimate the scale
uncertainty.
\begin{figure}[h]
  \begin{center}
    \includegraphics[width=16cm]{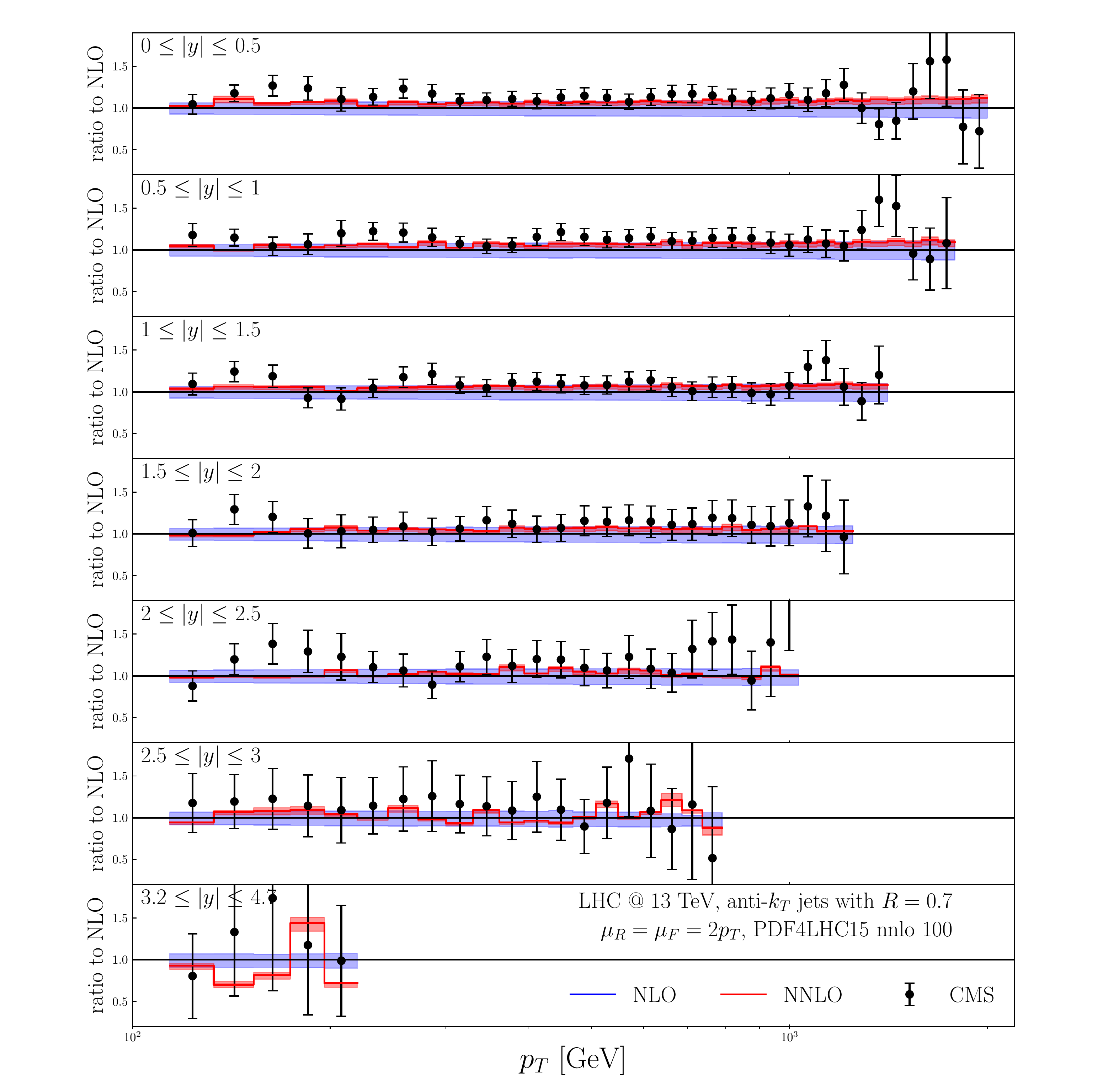}
  \end{center}
  \caption{\label{fig:inclusive-jet-pTxY} \sf Double-differential
    single jet inclusive cross-sections as measured by CMS
    \cite{Khachatryan:2016wdh} and NNLO perturbative QCD predictions
    as a function of the jet $p_T$ in slices of rapidity, for
    anti-$k_T$ jets with R = 0.7 normalised to the NLO result. Both
    perturbative predictions, NLO and NNLO, have been obtained with
    the PDF4LHC15\_nnlo PDF set and with $\mu_R = \mu_F = 2p_T$. The
    shaded bands represent the scale uncertainty obtained from
    differential distributions evaluated at $\mu_R = \mu_F = p_T$ and
    $\mu_R = \mu_F = 4p_T$.}
\end{figure}
In Fig.~\ref{fig:inclusive-jet-pTxY} we compare the NNLO QCD results
with the experimental measurement values from
Ref.~\cite{Khachatryan:2016wdh}. Results are normalised to the NLO QCD
prediction. The scale uncertainty of the latter is also shown. We do
not include non-perturbative or electroweak corrections. It is
expected that the non-perturbative effects are smaller for $R = 0.4$
than for $R = 0.7$. However, we are not interested in having the most
complete prediction, and thus choose somewhat arbitrarily one of the
$R$ values. Furthermore, with the current experimental precision,
neither non-perturbative nor electroweak effects are necessary to
obtain a good description of the data.
\begin{figure}[h]
  \begin{center}
    \includegraphics[width=11cm]{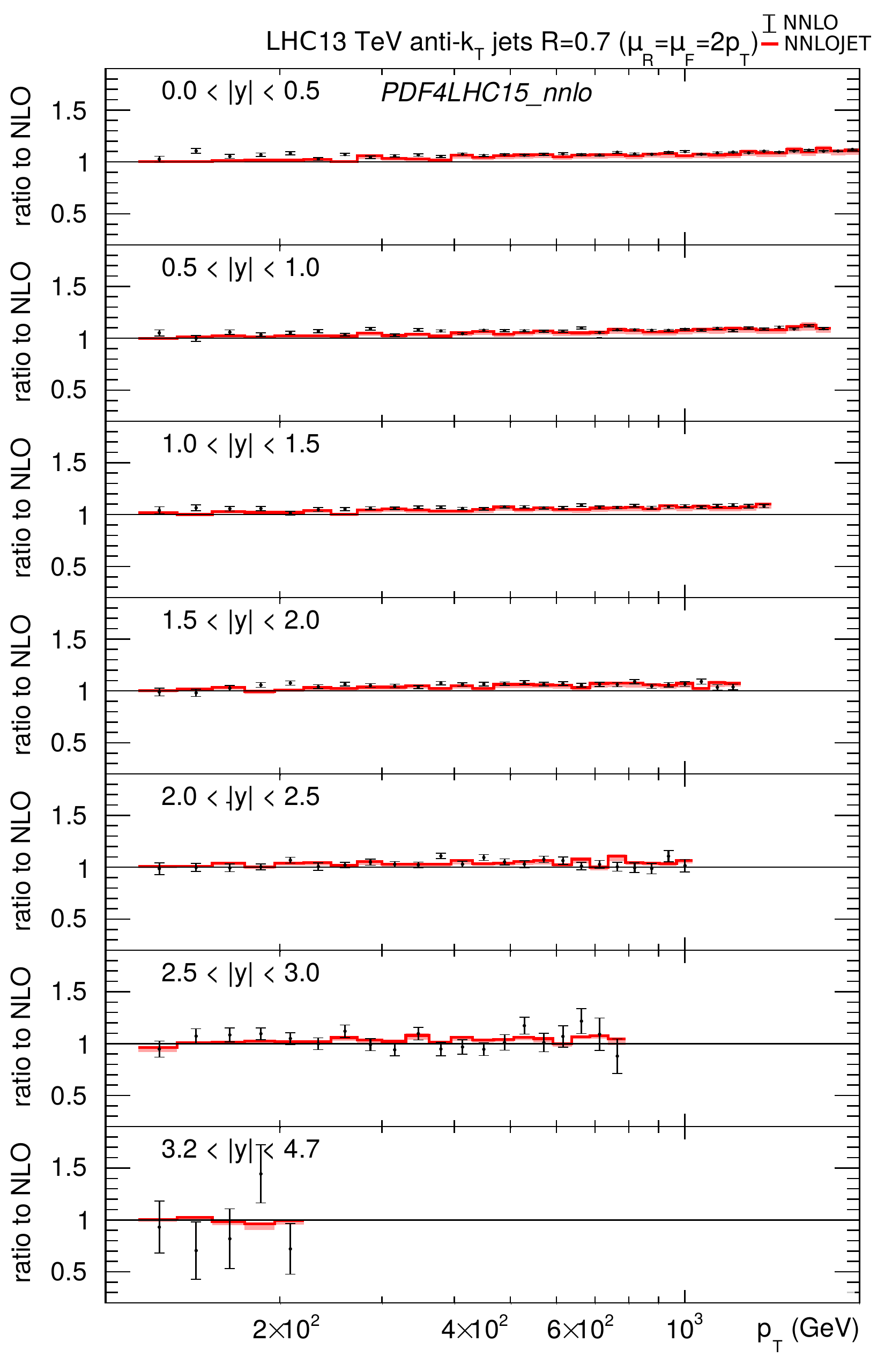}
  \end{center}
  \caption{\label{fig:comparison} \sf Comparison of the cross section
    ratios depicted in Fig.~\ref{fig:inclusive-jet-pTxY} as obtained
    with \textsc{NNLOjet} \cite{Currie:2018xkj} (red line with scale
    variation error, leading-color approximation for channels
    involving quarks) and with \textsc{Stripper} (black points with
    Monte Carlo integration error bars, as given in
    Appendix~\ref{app:k-factors}, exact in color). This figure has
    been obtained from Fig. 21 of \cite{Currie:2018xkj} by removing
    the experimental data points as well as the scale variation band
    of the NLO calculation, followed by superimposing the results
    obtained in the present work.}
\end{figure}
In Fig.~\ref{fig:comparison}, we compare our results with those
obtained with \textsc{NNLOjet} as presented in
Ref.~\cite{Currie:2018xkj}. Since no numerical values are given in the
latter publication, we superimpose our values, including the estimated
Monte Carlo integration error as listed in
Appendix~\ref{app:k-factors}, on the respective plot from
Ref.~\cite{Currie:2018xkj}. The results appear to be compatible within
their respective errors\footnote{The size of the integration errors of
the \textsc{NNLOjet} results can be judged from the fluctuations of
the K-factors}. The largest significant differences are observed in
the first rapidity slice at low jet transverse momenta. However, this
is the phase space region, where pure-gluon contributions
dominate. The latter have been compared separately (see
Section~\ref{sec:implementation}) and agree within one percent. We
also note that even though the bulk of the events are in the
low-$p_T$/central-rapidity region, our calculation is not optimised to
yield very small integration errors there. More interesting is the
comparison of the results for higher $p_T$ and in the first four
rapidity slices ($|y| < 2.0$). There, our calculation has an estimated
integration error at the level of about one percent, and is still
compatible with the \textsc{NNLOjet} result. This implies that
sub-leading color effects missing in the contributions of channels
involving quarks in \textsc{NNLOjet} are indeed at most at this
level. The integration errors of our calculation in the fifth rapidity
slice ($2.0 < |y| < 2.5$) are still less than about five percent and
the two calculations remain compatible, although \textsc{NNLOjet}
results are clearly more precise. While the integration errors in the
sixth rapidity slice ($2.5 < |y| < 3.0$) remain below ten percent, the
results can hardly be used as a precise indicator of sub-leading color
effects. We also provide the outcome of our calculation in the last
rapidity slice ($3.2 < |y| < 4.7$) to illustrate the limits of
reasonable convergence within our setup.

Let us finally comment on the convergence of the Monte Carlo
integration in \textsc{Stripper} for this process. The results
presented in Figs.~\ref{fig:inclusive-jet-pTxY} and
\ref{fig:comparison} required about 350000 CPU hours. In particular,
$\hat{\sigma}^{RR}_{F}$ was evaluated with 200000 CPU hours,
$\hat{\sigma}_{SU}$ with 100000 CPU hours and $\hat{\sigma}_{DU}$
with 50000 CPU hours. A further improvement of the integration errors
would require doubling the evaluation time for $\hat{\sigma}_{SU}$. It
is important to note that the computation cost of the integrated
subtraction terms present in $\hat{\sigma}_{DU}$ and
$\hat{\sigma}_{SU}$ is still less than that of the pure real
radiation. Hence, even if one could reduce it substantially by
performing analytic integrations of the subtraction terms, the
calculation would be at most twice faster for the same quality of the
results. To put the performance into perspective, we point out that
results for typical top-quark distributions as published recently
require less than a twentieth of the quoted running times.


\section{Outlook}

In the present publication, we have performed a first independent and
also the first complete calculation of single-jet inclusive rates for
LHC @ 13 TeV with NNLO QCD accuracy. After comparing with results
obtained with \textsc{NNLOjet}, we concluded that the sub-leading
color effects not included in \textsc{NNLOjet} are negligible for
phenomenological applications of the studied observable. One obvious
extension of the present work is to evaluate \textsc{fastNLO} and/or
\textsc{APPLGRID} tables for all measured jet observables in order to
allow for inclusion of the experimental data in PDF fits. In view of
the current computational costs, this requires either a substantial
improvement of the efficiency of \textsc{Stripper} or the acquisition
of substantial computational resources.

Apart from the calculation of jet rates, we have also discussed a
further evolution of the sector-improved residue subtraction scheme
which allows for a minimal number of subtraction terms per phase space
point. This approach does improve the convergence of cross sections,
but to quantify the improvement requires further studies. There are
still several avenues to explore in order to optimise the subtraction
scheme. They range from pure Monte Carlo techniques such as better
sampling of the initial state, through speed-ups of matrix element
evaluation by using analytic formulae, and finishing with further
modifications of the phase space treatment. We hope to be able to
substantially reduce the cost of calculations in the future.


\section*{Acknowledgements}

We would like to thank A.~Behring and D.~Heymes for their contributions
at early stages of this project, E.~W.~N.~Glover for providing
\textsc{NNLOjet} results for the pure-gluon case, and S.~Badger for
help with the \textsc{NJet} program.

A.M. thanks the Department of Physics at Princeton University for
hospitality during the completion of this work. The work of M.C. was
supported by the Deutsche Forschungsgemeinschaft under grant
396021762 - TRR 257. The research of A.M. and R.P. has received
funding from the European Research Council (ERC) under the European
Union's Horizon 2020 research and innovation programme (grant
agreement No 683211). The work of A.M. is also supported by the UK
STFC grants ST/L002760/1 and ST/K004883/1.
AvH was partly supported by grant No.\ 2015/17/B/ST2/01838 of the National
Science Center, Poland.

Simulations were performed with computing resources granted by RWTH
Aachen University under projects rwth0214, rwth0313 and rwth0414.

\paragraph{Note added.} After publication of the preprint of the present
work, we have received the cross section values obtained in
Ref.~\cite{Currie:2018xkj} with the help of \textsc{NNLOjet}. Our
results are statistically compatible with those values.


\appendix

\newpage

\section{Previous theoretical predictions for jet rates at NNLO in QCD}
\label{app:overview}

\paragraph{Single-jet inclusive cross sections}

\begin{enumerate}

\item Ref.~\cite{Currie:2016bfm} corresponding to 7 TeV ATLAS data
  presented in Ref.~\cite{Aad:2014vwa}
  \begin{itemize}
  \item jets defined with the anti-$k_T$ jet algorithm with $R = 0.4$,
    $p_T > 100$ GeV and $|y|<3.0$;
  \item hardest-jet scale: $\mu_R = \mu_F = \mu = p_{T,1}$.
  \end{itemize}

\item Ref.~\cite{Currie:2018xkj} corresponding to 13 TeV CMS data
  presented in Ref.~\cite{Khachatryan:2016wdh} (available on HEPDATA
  \url{https://www.hepdata.net/record/ins1459051})
  \begin{itemize}
  \item  jets defined with the anti-$k_T$ jet algorithm with $R=0.4$
    and $R = 0.7$, $p_T > 114$ GeV and $|y| < 4.7$;
  \item various scales: $\mu_R = \mu_F = \mu = p_{T,1}, \hat{H}_T,
    2p_T$.
  \end{itemize}

\end{enumerate}

\paragraph{Di-jet cross sections}

\begin{enumerate}

\item Ref.~\cite{Currie:2017eqf} corresponding to 7 TeV ATLAS data
  presented in Ref.~\cite{Aad:2013tea} (available on HEPDATA
  \url{https://www.hepdata.net/record/ins1268975})
  \begin{itemize}
  \item at least two jets identified with the anti-$k_T$ jet algorithm
    with $R=0.4$, $p_T > 100 (50)$ GeV for the leading (sub-leading) jet
    and $|y| < 3.0$;
  \item scale: $\mu = m_{jj}$ and $\mu = \langle p_T \rangle =
    (p_{T_1}+p_{T_2})/2$ (at NNLO both scales show similar behaviour);
  \item double-differential distributions ($m_{jj}$, $|y^*|$);
  \item includes electroweak corrections from
    Ref.~\cite{Dittmaier:2012kx}.
  \end{itemize}

\item Ref.~\cite{Gehrmann-DeRidder:2019ibf} corresponding to 8 TeV CMS data
  presented in Ref.~\cite{Sirunyan:2017skj} (available on
  HEPDATA \url{https://www.hepdata.net/record/ins1598460})
  \begin{itemize}
  \item at least two jets identified with the anti-$k_T$ jet
    algorithm with $R=0.7$, $p_T > 50$ GeV and $|y| < 3.0$;
  \item scale: $\mu = m_{jj}$;
  \item triple-differential distributions ($p_{T,\mathrm{avg}}$, $|y^*|$,
    $y_b$) in six $(y^*,y_b)$ regions (binning available from HEPDATA);
  \item includes non-perturbative and electroweak effects as
    multiplicative factors (bin-wise, available from HEPDATA).
  \end{itemize}

\end{enumerate}


\newpage

\section{Cross section contributions}
\label{app:notation}

At leading order
\be
\hat{\sigma}^{(0)}_{ab} = \hat{\sigma}^{{B}}_{ab} =
\f{1}{2\hat{s}} \f{1}{N_{ab}} \int \mathrm{d} \bm{\Phi}_n \, \la
\cm_n^{(0)} | \cm_n^{(0)} \ra \, F_n
\; ,
\ee
where $\hat{s} = (p_1+p_2)^2$ is the square of the partonic
center-of-mass energy, while $N_{ab}$ is the spin and color average
factor, defined as the product of the number of spin and color degrees
of freedom of the partons $a$ and $b$. The subscript $n$ points to the
number of final states in this contribution and $\mathrm{d}
\bm{\Phi}_n$ is the phase space measure for $n$
particles. $F_n$ is the infrared-safe measurement function
defining the observable. Here and below, $| \cm_n^{(l)} \ra$ are
$l$-loop amplitudes for $n$ particles understood as vectors in color
and spin space. At next-to-leading order there is
\be
\hat{\sigma}^{(1)}_{ab} = \hat{\sigma}^{{R}}_{ab} + 
\hat{\sigma}^{{V}}_{ab} + \hat{\sigma}^{{C}}_{ab} \; ,
\ee
with
\be
\begin{aligned} \label{eq:NLOXS}
\hat{\sigma}^{{R}}_{ab} &=
\f{1}{2\hat{s}} \f{1}{N_{ab}} \int \mathrm{d} \bm{\Phi}_{n+1} \, \la
\cm_{n+1}^{(0)} | \cm_{n+1}^{(0)} \ra \, F_{n+1} \; , \\[0.2cm]
\hat{\sigma}^{{V}}_{ab} &= \f{1}{2\hat{s}} \f{1}{N_{ab}} \int
\mathrm{d} \bm{\Phi}_n \, 2 \R \, \la \cm_n^{(0)} | \cm_n^{(1)} \ra \,
F_n \; , \\[0.2cm] 
\hat{\sigma}^{{C}}_{ab}(p_1,p_2) &= \f{\as}{2\pi} \f{1}{\ep}
\left( \f{\mR}{\mF} \right)^{\ep} \sum_c \int_0^1 \mathrm{d}z
\left[P^{(0)}_{ca}(z) \, \hat{\sigma}^{{B}}_{cb}(zp_1,p_2) +
  P^{(0)}_{cb}(z) \, \hat{\sigma}^{{B}}_{ac}(p_1,zp_2) \right]
\; ,
\end{aligned}
\ee
where $P^{(l)}_{ab}$ are Altarelli-Parisi splitting kernels at order $l$.
Finally, at next-to-next-to-leading order, we have
\be
\hat{\sigma}^{(2)}_{ab} = \hat{\sigma}^{{RR}}_{ab} +
\hat{\sigma}^{{RV}}_{ab} + \hat{\sigma}^{{VV}}_{ab} +
\hat{\sigma}^{{C1}}_{ab} + \hat{\sigma}^{{C2}}_{ab} \; ,
\ee
with
\be
\begin{aligned}
\hat{\sigma}^{{RR}}_{ab} &=
\f{1}{2\hat{s}} \f{1}{N_{ab}} \int \mathrm{d} \bm{\Phi}_{n+2}
\, \la \cm_{n+2}^{(0)} | \cm_{n+2}^{(0)} \ra \, F_{n+2} \; ,
\\[0.2cm] \hat{\sigma}^{{RV}}_{ab} &=
\f{1}{2\hat{s}} \f{1}{N_{ab}} \int \mathrm{d} \bm{\Phi}_{n+1} \, 2 \R
\, \la \cm_{n+1}^{(0)} | \cm_{n+1}^{(1)} \ra \, F_{n+1} \; ,
\\[0.2cm] \hat{\sigma}^{{VV}}_{ab} &=
\f{1}{2\hat{s}} \f{1}{N_{ab}} \int \mathrm{d} \bm{\Phi}_n \, \Big( 2
\R \, \la \cm_n^{(0)} | \cm_n^{(2)} \ra + \la \cm_n^{(1)} |
\cm_n^{(1)} \ra \Big) \, F_n \; ,
\end{aligned}
\ee
and
\be
\begin{split}
\label{eq:C1andC2}
\hat{\sigma}^{{C1}}_{ab}(p_1,p_2) &= \f{\as}{2\pi}
\f{1}{\ep} \left( \f{\mR}{\mF} \right)^{\ep} \sum_c \int_0^1\mathrm{d}z
\left[ P^{(0)}_{ca}(z) \, \hat{\sigma}^{{R}}_{cb}(zp_1,p_2) +
  P^{(0)}_{cb}(z)\, \hat{\sigma}^{{R}}_{ac}(p_1,zp_2) \right] \; ,
\\[0.2cm]
\hat{\sigma}^{{C2}}_{ab}(p_1,p_2) &=
\f{\as}{2\pi}\f{1}{\ep}
\left(\f{\mR}{\mF}\right)^{\ep} \sum_c \int_0^1\mathrm{d}z
\left[P^{(0)}_{ca}(z) \, \hat{\sigma}^{{V}}_{cb}(zp_1,p_2) +
  P^{(0)}_{cb}(z)\, \hat{\sigma}^{{V}}_{ac}(p_1,zp_2)
  \right] \\
&+\left(\f{\as}{2\pi}\right)^2\f{1}{2\ep}\left(\f{\mR}{\mF}\right)^{2\ep}
\sum_c \int_0^1\mathrm{d}z \left[P^{(1)}_{ca}(z)\,
  \hat{\sigma}^{{B}}_{cb}(zp_1,p_2) + P^{(1)}_{cb}(z)\,
  \hat{\sigma}^{{B}}_{ac}(p_1,zp_2)
  \right] \\
&+\left(\f{\as}{2\pi}\right)^2\f{\beta_0}{4\ep^2}
\left[\left(\f{\mR}{\mF}\right)^{2\ep}-2\left(\f{\mR}{\mF}\right)^{\ep}\right]
\sum_c \int_0^1\mathrm{d}z \left[P^{(0)}_{ca}(z)\,
  \hat{\sigma}^{{B}}_{cb}(zp_1,p_2) + P^{(0)}_{cb}(z)
  \,\hat{\sigma}^{{B}}_{ac}(p_1,zp_2)
  \right] \\
&+\left(\f{\as}{2\pi}\right)^2\f{1}{2\ep^2}\left(\f{\mR}{\mF}\right)^{2\ep}
\sum_{cd}\int_0^1\mathrm{d}z \left[\Big(P^{(0)}_{cd}\otimes P^{(0)}_{da}\Big)(z)\,
  \hat{\sigma}^{{B}}_{cb}(zp_1,p_2)  + \Big(P^{(0)}_{cd}\otimes
  P^{(0)}_{db}\Big)(z)\, \hat{\sigma}^{{B}}_{ac}(p_1,zp_2)
  \right] \\
&+\left(\f{\as}{2\pi}\right)^2\f{1}{\ep^2}\left(\f{\mR}{\mF}\right)^{2\ep}
\sum_{cd}\iint_0^1\mathrm{d}z\,\mathrm{d}\bar{z}
\left[P^{(0)}_{ca}(z)\,P^{(0)}_{db}({\bar z})\,
  \hat{\sigma}^{{B}}_{cd}(zp_1,\bar{z}p_2) 
    \right] \; ,
\end{split}
\ee
where
\be
\left(f \otimes g\right)(x)=
\iint_0^1\mathrm{d}y\,\mathrm{d}z\, f(y)g(z)\,\delta(x-yz) \; .
\ee
%


\newpage

\section{'t Hooft-Veltman corrections}
\label{app:tHVcor}

\subsection*{Double-real contributions}

The 't Hooft-Veltman corrections to double-pole contributions contained in
$\mathcal{S}_1$, $\mathcal{S}_4$, $\mathcal{S}_5$ and $\mathcal{S}_6$
(apart from the special case discussed below) are identical and given by
\be
  N^{RR}_0(0) \ni \int \dd{\mu}(u_1) \; 2 \ep
                h(\eta_1)^{-2\ep} \;
                \left(2 A_1 \ln (h(\eta_1)) +
                  (h(\eta_1)^{2\ep}-1)\left(A_2
                    +A_1\ln(\mu^2/E_\text{cms}^2)\right) \right) \; ,
\ee
together with the corresponding subtraction terms obtained by
expanding in $\eta_1$. Here, $h(\eta) =
E_\text{norm}/u^0_\text{1,max}(\eta)$ and
\begin{align}
  A_1 &= -\dd{\mu}^{(0)}(u_2) \; \dd\Phi_n \; \mathcal{S} \;
        \la \cm_{n+2}^{(0)} | \cm_{n+2}^{(0)} \ra
        \left(\frac{-\delta(\xi_1)}{4\ep}\right)
        \left(\frac{-\delta(\xi_2)}{2\ep}\right)
        \left(\frac{-\delta(\eta_2)}{a_{\eta_2}\ep}\right) \; , \\[0.2cm]
  A_2 &= -\dd {\mu}^{(1)}(u_2) \; \dd\Phi_n \; \mathcal{S} \;
        \la \cm_{n+2}^{(0)} | \cm_{n+2}^{(0)} \ra
        \left(\frac{-\delta(\xi_1)}{4\ep}\right)
        \left(\frac{-\delta(\xi_2)}{2\ep}\right)
        \left(\frac{-\delta(\eta_2)}{a_{\eta_2}\ep}\right) \; ,
\end{align}
where $\dd {\mu}(u_i) = \sum_{j=0} \dd {\mu}^{(j)}(u_i) \ep^j$ is the
integration measure for the unresolved parton momentum $u_i$,
$\mathcal{S}$ denotes the selector function, and $a_{\eta_2} = 1 $ for
$\mathcal{S}_{1,6}$ and $ a_{\eta_2}= 2$ for $ \mathcal{S}_{4,5}$. The
corrections to single-pole contributions depend on the sector and
can be written as
\be
 N^{RR}_0(0) \ni\int \dd{\mu}(u_i) \dd{\mu}^{(0)}(u_j) \;
      f^\text{tHV} \; \dd\Phi_n \; \mathcal{S} \; \la
      \cm_{n+2}^{(0)} | \cm_{n+2}^{(0)} \ra \prod_{i}
      \left(\frac{-\delta(x_i)}{a_{x_i}\ep}\right) \; ,
 \label{eq:tHVcorStructure}
\ee
where
\begin{center}
\begin{tabular}{c|c|c|c}
 $f^\text{tHV}$ & $x_i$ & $\dd{\mu}^{(0)}$ for & remarks \\
 \hline
 \multicolumn{4}{c}{$\mathcal{S}_1$ - $\eta_2$ and $\xi_2$ pole}\\
 \hline
 $2(h(\eta_1)^{-2\ep}-1)$ & $\xi_1,\xi_2$ & $u_2$ &
    $h(\eta) = E_\text{norm}/u^0_\text{1,max}(\eta)$\\
 $2(h(\eta_1)^{-2\ep}-1)$ & $\xi_1,\eta_2$ & $u_2$ &
    \dots\\
 \hline
 \multicolumn{4}{c}{$\mathcal{S}_{23}$ - $\eta_1$ and $\xi_2$ pole}\\
 \hline

 $(h(\xi_1)/\xi_1)^{-2\ep}-1$ &
  $\eta_1,\xi_2$ & $u_1$ &
    $h(\xi) = E_\text{norm}/u^0_\text{2,max}(\xi)$\\

 $2((h(0)/\xi_2)^{-2\ep}-1)$ &
  $\eta_1,\xi_1$ & $u_1$ &
    \dots\\

 $2\left(h(0)^{-2\ep}-1 +2\ep h(0)^{-2\ep}\ln(h(0)) \right) $ &
  $\eta_1,\xi_1,\xi_2$ & $u_1$ &
    \\

 $4 (h(0)^{-2\ep}-1)(1-2^{\ep}+\ep 2^{\ep}\ln(2))$ & $\eta_1,\xi_1,\eta_2,\xi_2$&
 $u_2$ & $h(\xi) = E_\text{norm}/u^0_\text{1,max}(\xi)$\\

 $2 (h(0)^{-2\ep}-1)(1-(2/\eta_2)^{-\ep})$ & $\eta_1,\xi_1,\xi_2$ &
 $u_2$ & \dots\\

 $2 (2^{\ep}-1-\ep 2^{\ep}\ln(2))$ & $\eta_1,\eta_2,\xi_2$ &

 $u_2$ &  \\
 $((2/\eta_2)^{-\ep}-1)$ & $\eta_1,\xi_2$ & $u_2$ & \\

 $4 (h(0)^{-2\ep}-1)(1-(2/\eta_1)^{-\ep})$ & $\xi_1,\eta_2,\xi_2$
  & $u_2$ & \\

 $2 (h(0)^{-2\ep}-1)$ & $\xi_1,\xi_2$ & $u_2$ & \\

 $2((2/\eta_1)^{-\ep}-1)$ & $\xi_1,\eta_2,\xi_2$ & $u_2$ & \\

 \hline
 \multicolumn{4}{c}{$\mathcal{S}_{4}$ - $\eta_2$ and $\xi_2$ pole}\\
 \hline

 $2(h(\eta_1)^{-2\ep}-1)$ & $\xi_1,\xi_2$ & $u_2$ &
    $h(\eta) = E_\text{norm}/u^0_\text{12}(\eta)$\\

 $2(h(\eta_1)^{-2\ep}-1)$ & $\xi_1,\eta_2$ & $u_2$ & \dots\\

 \hline
 \multicolumn{4}{c}{$\mathcal{S}_{5}$ - $\eta_1$ and $\xi_2$ pole}\\
 \hline

  $2((h(\eta_2)^{-2\ep}-1)$ & $\eta_1,\xi_1$ & $u_2$ &
  $h(\eta) = E_\text{norm}/u_{12}^0(\eta)$\\

  $4((h(\eta_1)^{-2\ep}-1)(1-(1-\eta_1/2)^{\ep})$ &
  $\xi_1,\eta_2,\xi_2$ & $u_2$ & \dots \\

 $2(h(\eta_1)^{-2\ep}-1)$ & $\xi_1,\xi_2$& $u_2$ & \\

 $2((1-\eta_1/2)^{\ep}-1)$ & $\eta_2,\xi_2$ & $u_2$ & \\

 \hline
 \multicolumn{4}{c}{$\mathcal{S}_{6}$ - $\eta_1,\eta_2$ and $\xi_2$ pole}\\
 \hline

 $(h(\xi_1)/\xi_1)^{-2\ep}-1$ &
  $\eta_1,\xi_2$ & $u_1$ &
    $h(\xi) = E_\text{norm}/u^0_\text{2,max}(\xi)$\\

 $2((h(0)/\xi_2)^{-2\ep}-1)$ &
  $\eta_1,\xi_1$ & $u_1$ &
    \dots\\

 $2\left(h(0)^{-2\ep}-1 + 2\ep h(0)^{-2\ep}\ln(h(0)) \right) $ &
  $\eta_1,\xi_1,\xi_2$ & $u_1$ &
    \\

 $2(h(\eta_1)^{-2\ep}-1)$ & $\xi_1,\xi_2$ & $u_2$ &
    $h(\eta) = E_\text{norm}/u^0_\text{1,max}(\eta)$\\

 $2(h(\eta_1)^{-2\ep}-1)$ & $\xi_1,\eta_2$ & $u_2$ &
    \dots\\
 \hline
\end{tabular}
\end{center}
and
\begin{center}
\begin{tabular}{c|c|c|c|c|c}
 Sector & $\mathcal{S}_1$  & $\mathcal{S}_{23}$ & $\mathcal{S}_{4}$&
 $\mathcal{S}_{5}$ & $\mathcal{S}_{6}$  \\
 \hline
 $\{a_{\eta_1},a_{\xi_1},a_{\eta_2},a_{\xi_2}\}$  & $\{2,4,1,2\}$ &
 $\{1,4,2,2\}$ & $\{2,4,2,2\}$   & $\{2,4,2,2\}$   & $\{1,4,1,2\}$
\end{tabular}
\end{center}

\paragraph{Special case of $\mathcal{S}_6$ with only four partons in the final
state.} (See section 4.3.2 of Ref.~\cite{Czakon:2014oma} for
details). Due to the modification of the angles and energies implied
by the boost from the partonic center-of-mass frame, more terms
contribute in this case. We define
\begin{align}
  h_{\xi,1} &= E_\text{norm}/u^0_\text{1,max,lab}\;,
  &h_{\xi,2} = E_\text{norm}/u^0_\text{2,max,lab}\;,\\[0.2cm]
  h_{\eta,1} &= \left[\frac{r^0_{1,\text{lab}}}{r^0_{1,\text{cms}}}
                \frac{u^0_{1,\text{lab}}}{u^0_{1,\text{cms}}}\right]\;,
  &h_{\eta,2} = \left[\frac{r^0_{2,\text{lab}}}{r^0_{2,\text{cms}}}
                \frac{u^0_{2,\text{lab}}}{u^0_{2,\text{cms}}}\right]\;.
\end{align}
The double-pole contribution gives rise to three different corrections
\begin{align}
  N^{RR}_0(0) \ni &\int \dd{\mu}(u_1)\; 2 \ep
                h_{\xi,1}^{-2\ep}
                \left(2 A_1 \ln \left(h_{\xi,1}\right) +
                  \left(h_{\xi,1}^{2\ep}-1\right)\left(A_2+
                  A_1\ep\ln\left(\mu^2/E_\text{cms}^2\right)\right)
                \right)
               \left(\frac{-\delta(\xi_1)}{4\ep}\right)\nn\\
               & \quad + \text{subtraction terms}\;,\\[0.2cm]
  N^{RR}_0(0) \ni &\int \dd{\mu}(u_1) \;
                  (h_{\eta,1}^{2\ep}-1)\left(A_1 +A_1\ep\ln(\mu^2/E_\text{cms}^2)+
                  A_2\ep\right)\left(\frac{-\delta(\eta_1)}{\ep}\right)\nn\\
               &\quad
               + \text{subtraction terms} \;,\\[0.2cm]
  N^{RR}_0(0) \ni &\int \dd{\mu}(u_1)\; 2 \ep
                h_{\xi,1}^{-2\ep}\left(1-h_{\eta,1}^{2\ep}\right) \left(2 A_1
               \ln (h_{\xi,1})
                +\left(h_{\xi,1}^{2\ep}-1\right) \left(A_2+
                 A_1\ep\ln(\mu^2/E_\text{cms}^2)\right)\right) \nn \\
                &\quad \times
               \left(\frac{-\delta(\eta_1)}{\ep}\right)
               \left(\frac{-\delta(\xi_1)}{4\ep}\right)
               + \text{subtraction terms}\;,
\end{align}
with
\begin{align}
  A_1 &= -\dd{\mu}^{(0)}(u_2)\;\dd\Phi_n\;\mathcal{S}\;
   \la\mathcal{M}_{n+2}^{(0)}|\mathcal{M}_{n+2}^{(0)}\ra
        \left(\frac{-\delta(\xi_2)}{2\ep}\right)
        \left(\frac{-\delta(\eta_2)}{\ep}\right)\;, \\[0.2cm]
  A_2 &= -\dd{\mu}^{(1)}(u_2)\;\dd\Phi_n\;\mathcal{S}\;
        \la\mathcal{M}_{n+2}^{(0)}|\mathcal{M}_{n+2}^{(0)}\ra
        \left(\frac{-\delta(\xi_2)}{2\ep}\right)
        \left(\frac{-\delta(\eta_2)}{\ep}\right) \;.
\end{align}
The single-pole contribution correction has the same structure as
\eqref{eq:tHVcorStructure} with
\begin{center}
\begin{tabular}{c|c|c}
 $f^\text{tHV}$ & $x_i$ & $\dd{\mu}^{(0)}$ for \\
 \hline
 \multicolumn{3}{c}{$\mathcal{S}_{6}$ - $\eta_1,\eta_2$ and $\xi_2$ pole,
                    special case}\\
 \hline
 $2(h_{\xi,1}^{-2\ep}-1)(1-h_{\eta,1}^{-\ep})$ &
   $\eta_1,\xi_1,\eta_2$ & $u_2$ \\
 $h_{\eta,1}^{-\ep}-1$ & $\eta_1,\eta_2$ & $u_2$  \\
 $2(h_{\xi,1}^{-2\ep}-1)$ & $\xi_1,\eta_2$ & $u_2$  \\

 $2(h_{\xi,1}^{-2\ep}-1)(1-h_{\eta,1}^{-\ep})$ &
  $\eta_1,\xi_1,\xi_2$ & $u_2$ \\
 $h_{\eta,1}^{-\ep}-1$ & $\eta_1,\xi_2$ & $u_2$ \\
 $2(h_{\xi,1}^{-2\ep}-1)$ & $\xi_1,\xi_2$ & $u_2$ \\

 $2(h_{\eta,2}^{-\ep}-1)(h_{\xi,2}^{-2\ep}-1 -2\ep
  h_{\xi,2}^{-2\ep}\ln(h_{\xi,2}))$ &
  $\eta_1,\xi_1,\eta_2,\xi_2$ & $u_1$ \\

 $2(h_{\eta,2}^{-\ep}-1)(1-(h_{\xi,2}/\xi_2)^{-2\ep})$ &
  $\eta_1,\xi_1,\eta_2$ & $u_1$ \\

 $2(-1+h_{\xi,2}^{-2\ep}-2\ep h_{\xi,2}^{-2\ep}\ln(h_{\xi,2}))$ &
  $\eta_1,\xi_1,\xi_2$ & $u_1$ \\

 $2((h_{\xi,2}/\xi_2)^{-2\ep}-1) $ & $\eta_1,\xi_1$ & $u_1$ \\

 $(h_{\eta,2}^{-\ep}-1)(1-(h_{\xi,2}/\xi_1)^{-2\ep})$ &
  $\eta_1,\xi_2,\eta_2$ & $u_1$ \\

 $(h_{\eta,2}^{-\ep}-1)$ &  $\eta_1,\eta_2$ & $u_1$ \\

 $(h_{\xi,2}/\xi_1)^{-2\ep}-1$ &
  $\eta_1,\xi_2$ & $u_1$ \\
 \hline
\end{tabular}
\end{center}

\subsection*{Real-virtual contributions}

The single-unresolved real-virtual contribution is given in
each sector by
\be
\begin{split}
  \sigh^{RV}_{SU} \ni &\frac{1}{2s}\frac{1}{N}\int \dd \Phi_{n+1}\;
  \mathcal{S}\;\left(2\Re\la\mathcal{M}_{n+1}^{(0)}|\bm{\mathrm{Z}}^{(1)}
  |\mathcal{M}_{n+1}^{(0)}\ra F_{n+1} + \text{subtraction terms}\right)\\[0.2cm]
  &\equiv \int \frac{\dd\eta}{\eta^{1-\ep}}\frac{\dd\xi}{\xi^{1-2\ep}}
  \left( f(\eta,\xi)+\text{subtraction terms}\right)
 \;.
\end{split}
\ee
Due to the virtual integrations, the scaling behaviour of
the $f(\eta,\xi)$ function is not trivial in the infrared limits
$\eta,\xi \to 0$. In the collinear limit there is
\begin{align}
  f(\eta,\xi) \xrightarrow{\eta\to 0}f^{(\eta,0)}(\xi)
  +\eta^{-\ep}\xi^{-\ep}f^{(\eta,1)}(\xi)
  +\eta^{-\ep}\xi^{-2\ep}f^{(\eta,2)}(\xi)\;,
\end{align}
while in the soft limit
\begin{align}
 f(\eta,\xi) \xrightarrow{\xi\to 0} f^{(\xi,0)}(\eta) + \xi^{-2\ep}f^{(\xi,1)}(\eta)
\;,
\end{align}
and finally in the soft-collinear limit
\begin{align}
   f(\eta,\xi) \xrightarrow{\eta\to 0,\;\xi \to 0} f^{(\eta\xi,0)} +
  \eta^{-\ep}\xi^{-2\ep}f^{(\eta\xi,1)}\;,
\end{align}
with $f^{(\eta\xi,0)}\equiv f^{(\eta,0)}(0)$ and $f^{(\eta\xi,1)}\equiv
f^{(\eta,2)}(0)$. This also implies that $f^{(\eta,1)}(0) = 0$. The
commutativity of soft and collinear limits implies for the soft
subtraction terms
\begin{align}
 f^{(\xi,1)}(\eta) \xrightarrow{\eta\to 0} \eta^{-\ep}f^{(\eta\xi,1)}\;.
\end{align}
Equivalently, we can define a function $f^{(\xi,\text{reg})} =
f^{(\xi,1)}(\eta) - \eta^{-\ep}f^{(\eta\xi,1)}$ which vanishes in the
$\eta\to 0$ limit.

For each sector contributing to $\sigh^{RV}_{SU}$ the following corrections are found.  For
the $f^{(c,0)}$ functions (ordinary scaling) there is
\be
\begin{split}
  N^{RV}_0(0) \ni  &\int \frac{\dd\eta}{\eta^{1-\ep}}\frac{-1}{2\ep}\left(
   \left(\frac{f^{(\xi,0)(-2)}(\eta)}{\ep^2} +
    \frac{f^{(\xi,0)(-1)}(\eta)}{\ep}\right)
   \left(h(\eta)^{-2\ep} -
   1\right)\right.\\[0.2cm]
   &\quad\left.+\left(\frac{f^{(\eta\xi,0)(-2)}}{\ep^2} +
   \frac{f^{(\eta\xi,0)(-1)}}{\ep}\right)
\left(h(0)^{-2\ep} - 1\right)\right)\;,
\end{split}
\ee
with $h(\eta) = E_\text{norm}/u^0_\text{max}(\eta)$ and the
$\ep$-expansions
\begin{align}
  f^{(c,n)} = \sum_{i=-2} f^{(c,n)(i)}\ep^i \; .
\end{align}
For the $f^{(\xi,1)}$ and $f^{(\eta\xi,1)}$ functions there is
\be
\begin{split}
  N^{RV}_0(0) \ni \int \frac{\dd\eta}{\eta^{1-\ep}}\frac{-1}{2\ep}&\left(\left(
   \frac{f^{(\xi,1)(-2)}(\eta)}{\ep^2}\right)
 \left(2\ep h(\eta)^{-2\ep}
\ln h(\eta)\right) \right.\\[0.2cm]
&-2\ep\left(\frac{f^{(\xi,1)(-1)}(\eta)}{\ep}\right)
\left(
  h(\eta)^{-2\ep} - 1\right)\\[0.2cm]
   &\quad+\left(\frac{f^{(\eta\xi,1)(-2)}(0)}{\ep^2}\right)
 \left(2\ep h(0)^{-2\ep}
\ln h(0)
+
\left(h(0)^{-2\ep} -
1\right)\ln\eta
\right) \\[0.2cm]
&\quad\left.-2\ep\left(\frac{f^{(\eta\xi,1)(-1)}(0)}{\ep}\right)
\left( h(0)^{-2\ep} - 1\right)\right)\;.
\end{split}
\ee
The following set of contributions involving the renormalisation scale
concludes the real-virtual contribution corrections
\begin{align}
  N^{RV}_0(0) \ni
  &  \int \frac{\dd\eta}{\eta^{1-\ep}}\frac{-1}{2\ep}\left(\left(
    \frac{f^{(\xi,1)(-2)}(\eta)}{\ep} \ln\left(\f{\mu_R^2}{E_\text{cms}^2}\right)\right)
    \left( h(\eta)^{-2\ep}-1 \right)\right.\nn\\[0.2cm]
  &\quad\left. +\left(\frac{f^{(\eta\xi,1)(-2)}}{\ep}\ln\left(\f{\mu_R^2}{E_\text{cms}^2}\right)\right)
    \left( h(0)^{-2\ep} - 1\right)\right)\;.
\end{align}

\subsection*{Collinear factorisation contributions}

The single-unresolved single-convolution contribution is given in each
sector (defined by the selector function $\mathcal{S}$) by
\be
\begin{split}
  \sigh^{C1}_{SU} \ni
 & \f{\as}{2\pi}
\f{1}{\ep} \left( \f{\mR}{\mF} \right)^{\ep} \sum_c \int_0^1\mathrm{d}z
\left( P^{(0)}_{ca}(z) \, \hat{\sigma}^{{R,\mathcal{S}}}_{cb}(zp_1,p_2) +
  P^{(0)}_{cb}(z)\, \hat{\sigma}^{{R,\mathcal{S}}}_{ac}(p_1,zp_2) \right)\\[0.2cm]
  &\equiv \frac{1}{\ep}\left(\f{\mR}{\mF}\right)^{\ep}
   \int \frac{\dd\eta}{\eta^{1-\ep}}\frac{\dd\xi}{\xi^{1-2\ep}}
 \left(f(\eta,\xi)+\text{subtraction terms}\right)
 \;.
\end{split}
\ee
After expanding the $\left(\mR/\mF\right)^{\ep}$ factor in $\ep$,
we find the following correction
\be
\begin{split}
N^{C1}_0 (0) \ni \int\frac{\dd\eta}{\eta^{1-\ep}}
\frac{-1}{2\ep}\left(\frac{f(\eta,0)}{\ep}
\left(h(\eta)^{-2\ep} - 1\right)
  - \frac{f(0,0)}{\ep}
\left(h(0)^{-2\ep} - 1\right)
\right)\;,
\end{split}
\ee
with $h(\eta) = E_\text{norm}/u^0_\text{max}(\eta)$.

\subsection*{Final remark}

Through the parameterised measurement function, a dependence on the
arbitrary energy scale $E_{\text{norm}}$ has been introduced. The
final result for the next-to-next-to-leading order cross section,
however, does not depend on this scale. The independence from
$E_{\text{norm}}$ can be used either as a check on the implementation
or to steer the cancellation of the arising logarithms.


\newpage

\section{Single-jet inclusive NNLO QCD K-factors}
\label{app:k-factors}

The following tables correspond to K-factors depicted in
Figs.~\ref{fig:inclusive-jet-pTxY}~and~\ref{fig:comparison}.
%
%
\begin{table}
\begin{tabularx}{\textwidth}{|C{0.85}|C{0.85}||C{1.3}|C{0.85}|C{1.1}|}
\hline
\multicolumn{5}{|C{5}|}{LHC @ 13 TeV, anti-$k_T$ jets with R =0.7,
  $\mu_R = \mu_F = 2 p_T$, PDF4LHC15\_nnlo} \\
\hline
\multicolumn{5}{|C{5}|}{$0.0 < |y| < 0.5$} \\
\hline
$p_{T,\mathrm{min}}$ [GeV] & $p_{T,\mathrm{max}}$ [GeV] &
$\sigma_{\mathrm{NLO}} $ [pb/GeV] &
$\sigma_{\mathrm{NNLO}}/\sigma_{\mathrm{NLO}}$ &
MC integration error [\%] \\
\hline \hline
114 & 133 & $7.842 \pm 0.016 \cdot 10^{3}$ & 1.026 & 2.81 \\
\hline
133 & 153 & $3.762 \pm 0.008 \cdot 10^{3}$ & 1.105 & 2.34 \\
\hline
153 & 174 & $1.89 \pm 0.004 \cdot 10^{3}$ & 1.053 & 2.03 \\
\hline
174 & 196 & $9.961 \pm 0.022 \cdot 10^{2}$ & 1.068 & 1.70 \\
\hline
196 & 220 & $5.398 \pm 0.011 \cdot 10^{2}$ & 1.082 & 1.52 \\
\hline
220 & 245 & $2.972 \pm 0.006 \cdot 10^{2}$ & 1.027 & 1.37 \\
\hline
245 & 272 & $1.692 \pm 0.003 \cdot 10^{2}$ & 1.071 & 1.22 \\
\hline
272 & 300 & $9.687 \pm 0.020 \cdot 10^{1}$ & 1.043 & 1.26 \\
\hline
300 & 330 & $5.692 \pm 0.012 \cdot 10^{1}$ & 1.058 & 1.14 \\
\hline
330 & 362 & $3.377 \pm 0.007 \cdot 10^{1}$ & 1.066 & 1.10 \\
\hline
362 & 395 & $2.036 \pm 0.004 \cdot 10^{1}$ & 1.052 & 1.01 \\
\hline
395 & 430 & $1.24 \pm 0.003 \cdot 10^{1}$ & 1.075 & 1.01 \\
\hline
430 & 468 & $7.578 \pm 0.015$ & 1.061 & 0.98 \\
\hline
468 & 507 & $4.695 \pm 0.010$ & 1.069 & 0.97 \\
\hline
507 & 548 & $2.94 \pm 0.006$ & 1.065 & 0.93 \\
\hline
548 & 592 & $1.842 \pm 0.004$ & 1.072 & 0.95 \\
\hline
592 & 638 & $1.162 \pm 0.003$ & 1.079 & 0.87 \\
\hline
638 & 686 & $7.387 \pm 0.015 \cdot 10^{-1}$ & 1.070 & 0.88 \\
\hline
686 & 737 & $4.697 \pm 0.009 \cdot 10^{-1}$ & 1.066 & 0.87 \\
\hline
737 & 790 & $2.992 \pm 0.006 \cdot 10^{-1}$ & 1.093 & 0.83 \\
\hline
790 & 846 & $1.93 \pm 0.004 \cdot 10^{-1}$ & 1.077 & 0.81 \\
\hline
846 & 905 & $1.236 \pm 0.002 \cdot 10^{-1}$ & 1.073 & 0.85 \\
\hline
905 & 967 & $7.931 \pm 0.016 \cdot 10^{-2}$ & 1.092 & 0.78 \\
\hline
967 & 1032 & $5.106 \pm 0.010 \cdot 10^{-2}$ & 1.100 & 0.79 \\
\hline
1032 & 1101 & $3.262 \pm 0.006 \cdot 10^{-2}$ & 1.074 & 0.74 \\
\hline
1101 & 1172 & $2.099 \pm 0.004 \cdot 10^{-2}$ & 1.087 & 0.78 \\
\hline
1172 & 1248 & $1.332 \pm 0.003 \cdot 10^{-2}$ & 1.094 & 0.74 \\
\hline
1248 & 1327 & $8.486 \pm 0.015 \cdot 10^{-3}$ & 1.087 & 0.76 \\
\hline
1327 & 1410 & $5.393 \pm 0.010 \cdot 10^{-3}$ & 1.103 & 0.77 \\
\hline
1410 & 1497 & $3.397 \pm 0.006 \cdot 10^{-3}$ & 1.093 & 0.68 \\
\hline
1497 & 1588 & $2.134 \pm 0.004 \cdot 10^{-3}$ & 1.103 & 0.69 \\
\hline
1588 & 1684 & $1.325 \pm 0.002 \cdot 10^{-3}$ & 1.112 & 0.72 \\
\hline
1684 & 1784 & $8.175 \pm 0.014 \cdot 10^{-4}$ & 1.102 & 0.70 \\
\hline
1784 & 1890 & $4.984 \pm 0.009 \cdot 10^{-4}$ & 1.105 & 0.70 \\
\hline
1890 & 2000 & $2.996 \pm 0.005 \cdot 10^{-4}$ & 1.120 & 0.67 \\
\hline
\end{tabularx}
\caption{NLO QCD single-inclusive cross section and NNLO QCD K-factors in the
         rapidity slice $0.0 < |y| < 0.5$.}
\end{table}


\begin{table}
\begin{tabularx}{\textwidth}{|C{0.85}|C{0.85}||C{1.3}|C{0.85}|C{1.1}|}
\hline
\multicolumn{5}{|C{5}|}{LHC @ 13 TeV, anti-$k_T$ jets with R =0.7,
  $\mu_R = \mu_F = 2 p_T$, PDF4LHC15\_nnlo} \\
\hline
\multicolumn{5}{|C{5}|}{$0.5 < |y| < 1.0$} \\
\hline
$p_{T,\mathrm{min}}$ [GeV] & $p_{T,\mathrm{max}}$ [GeV] &
$\sigma_{\mathrm{NLO}} $ [pb/GeV] &
$\sigma_{\mathrm{NNLO}}/\sigma_{\mathrm{NLO}}$ &
MC integration error [\%] \\
\hline \hline
114 & 133 & $7.423 \pm 0.016 \cdot 10^{3}$ & 1.051 & 2.89 \\
\hline
133 & 153 & $3.523 \pm 0.008 \cdot 10^{3}$ & 0.999 & 2.84 \\
\hline
153 & 174 & $1.775 \pm 0.004 \cdot 10^{3}$ & 1.058 & 2.23 \\
\hline
174 & 196 & $9.286 \pm 0.021 \cdot 10^{2}$ & 1.033 & 1.89 \\
\hline
196 & 220 & $5.012 \pm 0.011 \cdot 10^{2}$ & 1.052 & 1.60 \\
\hline
220 & 245 & $2.754 \pm 0.006 \cdot 10^{2}$ & 1.068 & 1.46 \\
\hline
245 & 272 & $1.549 \pm 0.003 \cdot 10^{2}$ & 1.033 & 1.30 \\
\hline
272 & 300 & $8.888 \pm 0.020 \cdot 10^{1}$ & 1.090 & 1.26 \\
\hline
300 & 330 & $5.206 \pm 0.012 \cdot 10^{1}$ & 1.029 & 1.28 \\
\hline
330 & 362 & $3.071 \pm 0.007 \cdot 10^{1}$ & 1.080 & 1.19 \\
\hline
362 & 395 & $1.837 \pm 0.004 \cdot 10^{1}$ & 1.070 & 1.09 \\
\hline
395 & 430 & $1.126 \pm 0.003 \cdot 10^{1}$ & 1.044 & 1.09 \\
\hline
430 & 468 & $6.821 \pm 0.016$ & 1.076 & 1.01 \\
\hline
468 & 507 & $4.184 \pm 0.010$ & 1.073 & 1.04 \\
\hline
507 & 548 & $2.605 \pm 0.006$ & 1.071 & 1.01 \\
\hline
548 & 592 & $1.63 \pm 0.004$ & 1.066 & 1.00 \\
\hline
592 & 638 & $1.011 \pm 0.002$ & 1.067 & 1.01 \\
\hline
638 & 686 & $6.443 \pm 0.015 \cdot 10^{-1}$ & 1.098 & 1.03 \\
\hline
686 & 737 & $4.066 \pm 0.010 \cdot 10^{-1}$ & 1.054 & 1.01 \\
\hline
737 & 790 & $2.568 \pm 0.006 \cdot 10^{-1}$ & 1.081 & 1.02 \\
\hline
790 & 846 & $1.638 \pm 0.004 \cdot 10^{-1}$ & 1.080 & 1.02 \\
\hline
846 & 905 & $1.038 \pm 0.002 \cdot 10^{-1}$ & 1.075 & 1.00 \\
\hline
905 & 967 & $6.589 \pm 0.017 \cdot 10^{-2}$ & 1.075 & 0.94 \\
\hline
967 & 1032 & $4.168 \pm 0.010 \cdot 10^{-2}$ & 1.085 & 0.93 \\
\hline
1032 & 1101 & $2.626 \pm 0.006 \cdot 10^{-2}$ & 1.077 & 1.02 \\
\hline
1101 & 1172 & $1.659 \pm 0.004 \cdot 10^{-2}$ & 1.094 & 1.17 \\
\hline
1172 & 1248 & $1.037 \pm 0.002 \cdot 10^{-2}$ & 1.073 & 0.96 \\
\hline
1248 & 1327 & $6.427 \pm 0.015 \cdot 10^{-3}$ & 1.098 & 1.03 \\
\hline
1327 & 1410 & $4.004 \pm 0.009 \cdot 10^{-3}$ & 1.088 & 0.95 \\
\hline
1410 & 1497 & $2.436 \pm 0.006 \cdot 10^{-3}$ & 1.106 & 0.97 \\
\hline
1497 & 1588 & $1.477 \pm 0.004 \cdot 10^{-3}$ & 1.089 & 0.93 \\
\hline
1588 & 1684 & $8.908 \pm 0.021 \cdot 10^{-4}$ & 1.119 & 0.96 \\
\hline
1684 & 1784 & $5.264 \pm 0.013 \cdot 10^{-4}$ & 1.092 & 0.99 \\
\hline
\end{tabularx}
\caption{NLO QCD single-inclusive cross section and NNLO QCD K-factors in the
         rapidity slice $0.5 < |y| < 1.0$.}
\end{table}


\begin{table}
\begin{tabularx}{\textwidth}{|C{0.85}|C{0.85}||C{1.3}|C{0.85}|C{1.1}|}
\hline
\multicolumn{5}{|C{5}|}{LHC @ 13 TeV, anti-$k_T$ jets with R =0.7,
  $\mu_R = \mu_F = 2 p_T$, PDF4LHC15\_nnlo} \\
\hline
\multicolumn{5}{|C{5}|}{$1.0 < |y| < 1.5$} \\
\hline
$p_{T,\mathrm{min}}$ [GeV] & $p_{T,\mathrm{max}}$ [GeV] &
$\sigma_{\mathrm{NLO}} $ [pb/GeV] &
$\sigma_{\mathrm{NNLO}}/\sigma_{\mathrm{NLO}}$ &
MC integration error [\%] \\
\hline \hline
114 & 133 & $6.553 \pm 0.016 \cdot 10^{3}$ & 1.039 & 3.55 \\
\hline
133 & 153 & $3.092 \pm 0.007 \cdot 10^{3}$ & 1.064 & 2.77 \\
\hline
153 & 174 & $1.539 \pm 0.004 \cdot 10^{3}$ & 1.056 & 2.17 \\
\hline
174 & 196 & $8.037 \pm 0.021 \cdot 10^{2}$ & 1.057 & 1.97 \\
\hline
196 & 220 & $4.289 \pm 0.011 \cdot 10^{2}$ & 1.013 & 1.85 \\
\hline
220 & 245 & $2.342 \pm 0.006 \cdot 10^{2}$ & 1.047 & 1.91 \\
\hline
245 & 272 & $1.309 \pm 0.003 \cdot 10^{2}$ & 1.052 & 1.54 \\
\hline
272 & 300 & $7.43 \pm 0.019 \cdot 10^{1}$ & 1.061 & 1.41 \\
\hline
300 & 330 & $4.308 \pm 0.011 \cdot 10^{1}$ & 1.059 & 1.33 \\
\hline
330 & 362 & $2.507 \pm 0.007 \cdot 10^{1}$ & 1.070 & 1.33 \\
\hline
362 & 395 & $1.496 \pm 0.004 \cdot 10^{1}$ & 1.068 & 1.29 \\
\hline
395 & 430 & $8.984 \pm 0.024$ & 1.058 & 1.25 \\
\hline
430 & 468 & $5.399 \pm 0.014$ & 1.054 & 1.22 \\
\hline
468 & 507 & $3.287 \pm 0.009$ & 1.073 & 1.25 \\
\hline
507 & 548 & $2.015 \pm 0.006$ & 1.073 & 1.19 \\
\hline
548 & 592 & $1.236 \pm 0.004$ & 1.063 & 1.20 \\
\hline
592 & 638 & $7.565 \pm 0.022 \cdot 10^{-1}$ & 1.063 & 1.22 \\
\hline
638 & 686 & $4.677 \pm 0.014 \cdot 10^{-1}$ & 1.089 & 1.33 \\
\hline
686 & 737 & $2.892 \pm 0.009 \cdot 10^{-1}$ & 1.068 & 1.42 \\
\hline
737 & 790 & $1.779 \pm 0.005 \cdot 10^{-1}$ & 1.067 & 1.35 \\
\hline
790 & 846 & $1.095 \pm 0.003 \cdot 10^{-1}$ & 1.084 & 1.30 \\
\hline
846 & 905 & $6.727 \pm 0.021 \cdot 10^{-2}$ & 1.065 & 1.41 \\
\hline
905 & 967 & $4.084 \pm 0.013 \cdot 10^{-2}$ & 1.072 & 1.30 \\
\hline
967 & 1032 & $2.483 \pm 0.008 \cdot 10^{-2}$ & 1.081 & 1.45 \\
\hline
1032 & 1101 & $1.49 \pm 0.005 \cdot 10^{-2}$ & 1.070 & 1.35 \\
\hline
1101 & 1172 & $8.856 \pm 0.029 \cdot 10^{-3}$ & 1.082 & 1.57 \\
\hline
1172 & 1248 & $5.179 \pm 0.017 \cdot 10^{-3}$ & 1.091 & 1.44 \\
\hline
1248 & 1327 & $3.004 \pm 0.010 \cdot 10^{-3}$ & 1.080 & 1.75 \\
\hline
1327 & 1410 & $1.698 \pm 0.006 \cdot 10^{-3}$ & 1.080 & 1.61 \\
\hline
\end{tabularx}
\caption{NLO QCD single-inclusive cross section and NNLO QCD K-factors in the
         rapidity slice $1.0 < |y| < 1.5$.}
\end{table}


\begin{table}
\begin{tabularx}{\textwidth}{|C{0.85}|C{0.85}||C{1.3}|C{0.85}|C{1.1}|}
\hline
\multicolumn{5}{|C{5}|}{LHC @ 13 TeV, anti-$k_T$ jets with R =0.7,
  $\mu_R = \mu_F = 2 p_T$, PDF4LHC15\_nnlo} \\
\hline
\multicolumn{5}{|C{5}|}{$1.5 < |y| < 2.0$} \\
\hline
$p_{T,\mathrm{min}}$ [GeV] & $p_{T,\mathrm{max}}$ [GeV] &
$\sigma_{\mathrm{NLO}} $ [pb/GeV] &
$\sigma_{\mathrm{NNLO}}/\sigma_{\mathrm{NLO}}$ &
MC integration error [\%] \\
\hline \hline
114 & 133 & $5.412 \pm 0.015 \cdot 10^{3}$ & 0.987 & 3.69 \\
\hline
133 & 153 & $2.516 \pm 0.007 \cdot 10^{3}$ & 0.980 & 3.40 \\
\hline
153 & 174 & $1.237 \pm 0.004 \cdot 10^{3}$ & 1.024 & 2.82 \\
\hline
174 & 196 & $6.35 \pm 0.021 \cdot 10^{2}$ & 1.056 & 2.30 \\
\hline
196 & 220 & $3.349 \pm 0.010 \cdot 10^{2}$ & 1.075 & 2.06 \\
\hline
220 & 245 & $1.813 \pm 0.006 \cdot 10^{2}$ & 1.039 & 2.06 \\
\hline
245 & 272 & $9.919 \pm 0.032 \cdot 10^{1}$ & 1.063 & 1.82 \\
\hline
272 & 300 & $5.527 \pm 0.018 \cdot 10^{1}$ & 1.051 & 1.77 \\
\hline
300 & 330 & $3.169 \pm 0.011 \cdot 10^{1}$ & 1.046 & 1.78 \\
\hline
330 & 362 & $1.804 \pm 0.006 \cdot 10^{1}$ & 1.035 & 1.63 \\
\hline
362 & 395 & $1.054 \pm 0.004 \cdot 10^{1}$ & 1.074 & 1.61 \\
\hline
395 & 430 & $6.133 \pm 0.023$ & 1.061 & 1.68 \\
\hline
430 & 468 & $3.586 \pm 0.014$ & 1.064 & 1.71 \\
\hline
468 & 507 & $2.094 \pm 0.008$ & 1.067 & 1.77 \\
\hline
507 & 548 & $1.238 \pm 0.005$ & 1.079 & 1.81 \\
\hline
548 & 592 & $7.211 \pm 0.034 \cdot 10^{-1}$ & 1.063 & 1.80 \\
\hline
592 & 638 & $4.25 \pm 0.024 \cdot 10^{-1}$ & 1.068 & 1.82 \\
\hline
638 & 686 & $2.453 \pm 0.011 \cdot 10^{-1}$ & 1.054 & 1.87 \\
\hline
686 & 737 & $1.414 \pm 0.007 \cdot 10^{-1}$ & 1.062 & 2.02 \\
\hline
737 & 790 & $8.047 \pm 0.039 \cdot 10^{-2}$ & 1.059 & 2.07 \\
\hline
790 & 846 & $4.524 \pm 0.023 \cdot 10^{-2}$ & 1.087 & 2.05 \\
\hline
846 & 905 & $2.525 \pm 0.013 \cdot 10^{-2}$ & 1.043 & 2.03 \\
\hline
905 & 967 & $1.385 \pm 0.008 \cdot 10^{-2}$ & 1.058 & 2.15 \\
\hline
967 & 1032 & $7.397 \pm 0.043 \cdot 10^{-3}$ & 1.066 & 2.26 \\
\hline
1032 & 1101 & $3.812 \pm 0.022 \cdot 10^{-3}$ & 1.089 & 2.53 \\
\hline
1101 & 1172 & $1.961 \pm 0.012 \cdot 10^{-3}$ & 1.033 & 2.58 \\
\hline
1172 & 1248 & $9.821 \pm 0.063 \cdot 10^{-4}$ & 1.036 & 2.75 \\
\hline
\end{tabularx}
\caption{NLO QCD single-inclusive cross section and NNLO QCD K-factors in the
         rapidity slice $1.5 < |y| < 2.0$.}
\end{table}


\begin{table}
\begin{tabularx}{\textwidth}{|C{0.85}|C{0.85}||C{1.3}|C{0.85}|C{1.1}|}
\hline
\multicolumn{5}{|C{5}|}{LHC @ 13 TeV, anti-$k_T$ jets with R =0.7,
  $\mu_R = \mu_F = 2 p_T$, PDF4LHC15\_nnlo} \\
\hline
\multicolumn{5}{|C{5}|}{$2.0 < |y| < 2.5$} \\
\hline
$p_{T,\mathrm{min}}$ [GeV] & $p_{T,\mathrm{max}}$ [GeV] &
$\sigma_{\mathrm{NLO}} $ [pb/GeV] &
$\sigma_{\mathrm{NNLO}}/\sigma_{\mathrm{NLO}}$ &
MC integration error [\%] \\
\hline \hline
114 & 133 & $4.118 \pm 0.015 \cdot 10^{3}$ & 0.984 & 5.67 \\
\hline
133 & 153 & $1.884 \pm 0.007 \cdot 10^{3}$ & 0.996 & 3.93 \\
\hline
153 & 174 & $9.057 \pm 0.035 \cdot 10^{2}$ & 0.991 & 3.71 \\
\hline
174 & 196 & $4.552 \pm 0.019 \cdot 10^{2}$ & 1.002 & 2.89 \\
\hline
196 & 220 & $2.331 \pm 0.010 \cdot 10^{2}$ & 1.066 & 2.84 \\
\hline
220 & 245 & $1.22 \pm 0.006 \cdot 10^{2}$ & 1.004 & 3.60 \\
\hline
245 & 272 & $6.481 \pm 0.029 \cdot 10^{1}$ & 1.019 & 2.72 \\
\hline
272 & 300 & $3.498 \pm 0.017 \cdot 10^{1}$ & 1.047 & 2.94 \\
\hline
300 & 330 & $1.893 \pm 0.010 \cdot 10^{1}$ & 1.025 & 2.80 \\
\hline
330 & 362 & $1.033 \pm 0.005 \cdot 10^{1}$ & 1.020 & 2.86 \\
\hline
362 & 395 & $5.608 \pm 0.032$ & 1.106 & 2.60 \\
\hline
395 & 430 & $3.038 \pm 0.018$ & 1.027 & 2.75 \\
\hline
430 & 468 & $1.621 \pm 0.010$ & 1.092 & 2.92 \\
\hline
468 & 507 & $8.719 \pm 0.059 \cdot 10^{-1}$ & 1.050 & 2.99 \\
\hline
507 & 548 & $4.59 \pm 0.032 \cdot 10^{-1}$ & 1.027 & 3.23 \\
\hline
548 & 592 & $2.383 \pm 0.019 \cdot 10^{-1}$ & 1.074 & 3.16 \\
\hline
592 & 638 & $1.21 \pm 0.011 \cdot 10^{-1}$ & 1.062 & 3.59 \\
\hline
638 & 686 & $6.044 \pm 0.052 \cdot 10^{-2}$ & 1.009 & 3.57 \\
\hline
686 & 737 & $2.878 \pm 0.025 \cdot 10^{-2}$ & 1.025 & 3.81 \\
\hline
737 & 790 & $1.346 \pm 0.014 \cdot 10^{-2}$ & 1.003 & 4.19 \\
\hline
790 & 846 & $6.103 \pm 0.065 \cdot 10^{-3}$ & 0.993 & 4.57 \\
\hline
846 & 905 & $2.681 \pm 0.032 \cdot 10^{-3}$ & 0.985 & 5.10 \\
\hline
905 & 967 & $1.071 \pm 0.015 \cdot 10^{-3}$ & 1.105 & 5.43 \\
\hline
967 & 1032 & $4.172 \pm 0.062 \cdot 10^{-4}$ & 1.014 & 6.02 \\
\hline
\end{tabularx}
\caption{NLO QCD single-inclusive cross section and NNLO QCD K-factors in the
         rapidity slice $2.0 < |y| < 2.5$.}
\end{table}


\begin{table}
\begin{tabularx}{\textwidth}{|C{0.85}|C{0.85}||C{1.3}|C{0.85}|C{1.1}|}
\hline
\multicolumn{5}{|C{5}|}{LHC @ 13 TeV, anti-$k_T$ jets with R =0.7,
  $\mu_R = \mu_F = 2 p_T$, PDF4LHC15\_nnlo} \\
\hline
\multicolumn{5}{|C{5}|}{$2.5 < |y| < 3.0$} \\
\hline
$p_{T,\mathrm{min}}$ [GeV] & $p_{T,\mathrm{max}}$ [GeV] &
$\sigma_{\mathrm{NLO}} $ [pb/GeV] &
$\sigma_{\mathrm{NNLO}}/\sigma_{\mathrm{NLO}}$ &
MC integration error [\%] \\
\hline \hline
114 & 133 & $2.817 \pm 0.018 \cdot 10^{3}$ & 0.944 & 7.76 \\
\hline
133 & 153 & $1.238 \pm 0.007 \cdot 10^{3}$ & 1.070 & 7.15 \\
\hline
153 & 174 & $5.707 \pm 0.033 \cdot 10^{2}$ & 1.081 & 6.68 \\
\hline
174 & 196 & $2.733 \pm 0.018 \cdot 10^{2}$ & 1.093 & 5.66 \\
\hline
196 & 220 & $1.321 \pm 0.009 \cdot 10^{2}$ & 1.046 & 5.70 \\
\hline
220 & 245 & $6.386 \pm 0.050 \cdot 10^{1}$ & 0.996 & 5.63 \\
\hline
245 & 272 & $3.114 \pm 0.023 \cdot 10^{1}$ & 1.117 & 5.98 \\
\hline
272 & 300 & $1.473 \pm 0.012 \cdot 10^{1}$ & 0.987 & 5.85 \\
\hline
300 & 330 & $7.126 \pm 0.066$ & 0.938 & 5.88 \\
\hline
330 & 362 & $3.352 \pm 0.033$ & 1.094 & 6.07 \\
\hline
362 & 395 & $1.55 \pm 0.016$ & 0.945 & 6.52 \\
\hline
395 & 430 & $7.04 \pm 0.083 \cdot 10^{-1}$ & 0.965 & 6.87 \\
\hline
430 & 468 & $3.151 \pm 0.040 \cdot 10^{-1}$ & 0.942 & 6.09 \\
\hline
468 & 507 & $1.263 \pm 0.019 \cdot 10^{-1}$ & 1.009 & 7.45 \\
\hline
507 & 548 & $4.999 \pm 0.079 \cdot 10^{-2}$ & 1.170 & 8.08 \\
\hline
548 & 592 & $1.906 \pm 0.035 \cdot 10^{-2}$ & 1.008 & 8.94 \\
\hline
592 & 638 & $6.703 \pm 0.138 \cdot 10^{-3}$ & 1.067 & 10.36 \\
\hline
638 & 686 & $2.084 \pm 0.052 \cdot 10^{-3}$ & 1.214 & 12.06 \\
\hline
686 & 737 & $5.811 \pm 0.250 \cdot 10^{-4}$ & 1.088 & 15.55 \\
\hline
737 & 790 & $1.558 \pm 0.059 \cdot 10^{-4}$ & 0.877 & 16.64 \\
\hline
\end{tabularx}
\caption{NLO QCD single-inclusive cross section and NNLO QCD K-factors in the
         rapidity slice $2.5 < |y| < 3.0$.}
\end{table}


\begin{table}
\begin{tabularx}{\textwidth}{|C{0.85}|C{0.85}||C{1.3}|C{0.85}|C{1.1}|}
\hline
\multicolumn{5}{|C{5}|}{LHC @ 13 TeV, anti-$k_T$ jets with R =0.7,
  $\mu_R = \mu_F = 2 p_T$, PDF4LHC15\_nnlo} \\
\hline
\multicolumn{5}{|C{5}|}{$3.2 < |y| < 4.7$} \\
\hline
$p_{T,\mathrm{min}}$ [GeV] & $p_{T,\mathrm{max}}$ [GeV] &
$\sigma_{\mathrm{NLO}} $ [pb/GeV] &
$\sigma_{\mathrm{NNLO}}/\sigma_{\mathrm{NLO}}$ &
MC integration error [\%] \\
\hline \hline
114 & 133 & $4.808 \pm 0.056 \cdot 10^{2}$ & 0.927 & 25.06 \\
\hline
133 & 153 & $1.646 \pm 0.018 \cdot 10^{2}$ & 0.701 & 27.72 \\
\hline
153 & 174 & $5.676 \pm 0.092 \cdot 10^{1}$ & 0.815 & 28.80 \\
\hline
174 & 196 & $2.109 \pm 0.039 \cdot 10^{1}$ & 1.441 & 28.08 \\
\hline
196 & 220 & $7.464 \pm 0.117$ & 0.718 & 24.45 \\
\hline
\end{tabularx}
\caption{NLO QCD single-inclusive cross section and NNLO QCD K-factors in the
         rapidity slice $3.2 < |y| < 4.7$.}
\end{table}


\newpage

\bibliographystyle{JHEP}
\bibliography{references}

\end{document}